\documentclass[11pt]{article}

\usepackage[utf8]{inputenc} 
\usepackage[T1]{fontenc}    
\usepackage{csquotes}
\usepackage{hyperref}       
\usepackage{url}            
\usepackage{amsfonts}       
\usepackage{textcomp}

\usepackage{graphicx}

\usepackage{algorithm}
\usepackage{algpseudocode}

\usepackage{amsmath}
\usepackage{bm}
\usepackage{amssymb}
\usepackage{amsthm}

\usepackage{booktabs}
\usepackage{multirow}

\usepackage[a4paper, left=3cm, right=2.5cm, top=3cm, bottom=3cm]{geometry}

\usepackage{tikz}
\usetikzlibrary{arrows,shapes.geometric}
\usepackage{tikz-qtree}
\usepackage{tango-colors}

\usepackage{authblk}
\usepackage{natbib}
\usepackage{doi}

\newcommand{\conspace}{\mathcal{X}}

\begin{document}

\title{Automatic Creativity Measurement in Scratch Programs Across Modalities}

\author[1]{Anastasia~Kovalkov}
\author[2,3]{Benjamin~Paaßen}
\author[1]{Avi~Segal}
\author[2,3]{Niels~Pinkwart}
\author[1,4]{Kobi~Gal}
\affil[1]{Ben-Gurion University of the Negev}
\affil[2]{Humboldt University of Berlin}
\affil[3]{German Research Center for Artificial Intelligence}
\affil[4]{University of Edinburgh}

\date{\copyright 2021 IEEE. Personal use of this material is permitted.  Permission from IEEE must be obtained for all other uses, in any current or future media, including reprinting/republishing this material for advertising or promotional purposes, creating new collective works, for resale or redistribution to servers or lists, or reuse of any copyrighted component of this work in other works.} 
\pagestyle{myheadings}
\markright{\copyright 2021 IEEE. Refer to \href{https://doi.org/10.1109/TLT.2022.3144442}{doi:10.1109/TLT.2022.3144442} for original article.}

\maketitle

\begin{abstract}
Promoting creativity is considered an important goal of education, but creativity is notoriously hard
to measure. 
In this paper, we make the journey from  defining a formal measure of creativity that is efficiently 
computable to applying the measure in a practical domain. The measure is general and relies on core 
theoretical concepts in creativity theory, namely fluency, flexibility, and originality,
integrating  with prior cognitive science literature.
We adapted the general measure for projects in the popular visual programming language Scratch. 
We designed a machine learning model for predicting the creativity of Scratch projects, trained and
evaluated on human expert creativity assessments in an extensive user study.
Our results show that opinions about creativity in Scratch varied widely across experts. The automatic creativity assessment aligned with the assessment of the human experts more than the experts agreed with each other. This is a first step in providing computational models for measuring creativity that can be applied to educational technologies, and to scale up the benefit of creativity education in schools.
\end{abstract}


\section{Introduction}

Creativity has been shown to promote students’ critical thinking, self-motivation,
and mastery of skills and concepts \cite{Henriksen2016,Knobelsdorf2008,Resnick2017}.
To track student's creative achievement, we require instruments to quantify creativity.
However, creativity is notoriously hard to quantify because it is
highly context-dependent \cite{Amabile2018,Henriksen2016}.
\cite{Amabile2018} addressed this challenge by relying on a panel of domain experts who
assess each creative product. Unfortunately, it is hardly
feasible to ask an expert panel to rate the creative products of millions of students.
This begs the question whether it is possible to construct an automated creativity assessment
technique that can serve as a surrogate for expert assessment.

Prior attempts at automating creativity assessment include \cite{Huang2010,Kovalkov2020,Yeh2015}.
All these approaches build upon creativity tests from the psychometric literature 
\cite{Torrance1972,Williams1979,Runco2016}, in particular the fluency, flexibility,
and originality scales from Torrance's test of creative thinking \cite{Torrance1972}.
Fluency refers to the number of generated ideas, flexibility to the number of distinct classes of 
generated ideas, and originality to the infrequency of ideas compared to a typical population.
Prior approaches have implemented automatic versions of these scales for specific domains
\cite{Huang2010,Kovalkov2020,Yeh2015}.

In this paper, we propose a novel formalization of fluency, flexibility,
and originality that is flexible enough to be adapted across
modalities while remaining efficient to compute. Our formalization requires
two ingredients: A set of concepts for each modality, and a distance between these
concepts. Then, we can treat any creative product as a structured combination
of concepts from each modality. We measure fluency as the distance
to an empty product, flexibility as the distance between concepts in the product, and
originality as the average distance to typical products.
We show that our measure is a proper generalization of Torrance's scales \cite{Torrance1972}, that is,
we obtain Torrance's notions of fluency, flexibility, and originality as a special case for
a certain distance between concepts.

Note that our formalization is designed to be general. To demonstrate its practical feasibility,
we implement it for a specific multimodal domain, namely Scratch, a visual programming
environment designed for open-ended, creative learning \cite{Maloney2010}. In doing so, we expand
our prior work which considered only code and images \cite{Kovalkov2021}. Scratch is particularly
interesting for automated assessment because many Scratch students program on their own without access
to teacher feedback, such that an automatic feedback system would benefit them in particular.

We conducted a user study to collect expert assessments of creativity of Scratch projects.
The experts were Scratch instructors without prior knowledge in creativity theory. 
Each expert was assigned a set of preselected Scratch projects and asked to separately assess 
the creativity of projects according to four different aspects: code, visuals, audio, and idea behind
the project. 

We designed an online application to facilitate the assessment process, which allowed the experts to
interact with each project as needed. 
We compared the resulting expert assessments to our automatic assessments. We find that the
automatic assessments agree with expert assessments at least as much as experts agree with each other.

The contribution of this work is twofold. First, we provide a novel and general computational
framework to measure creativity, which extends classic notions of creativity in the literature.
Second, we implement this framework for Scratch and thus provide an automatic measure to scale up
creativity assessment.

\section{Related Work}

The largest body of prior work on creativity stems from psychology, where creativity
is considered a crucial aspect of the human intellect \cite{Runco2012}. 
Guilford's model on the structure of intellect includes creativity as
\emph{divergent production}, meaning the ability to generate a wide variety of ideas
on the same topic \cite{Guilford1956}.
Based on this definition of creativity, creativity tests have been developed, which
present participants with a prompt and ask for as many ideas as possible
in reaction to that prompt \cite{Torrance1972,Williams1979,Runco2016,Kim2006}.

Consider the example in Fig.~\ref{fig:creativity_example}.
Here, the prompt consists of three geometric shapes and the task
is to generate as many objects from these shapes as possible.
In Torrance's test of creative thinking \cite{Torrance1972}, we
then count the number of generated objects and call this number \emph{fluency};
we call the number of distinct classes of objects \emph{flexibility},
and we call the infrequency of objects compared to a typical population of
participants \emph{originality} \cite{Torrance1972}.
In this work, we focus particularly on these three scales from Torrance's
test as prior research has already established connections between
Torrance's scales and success in beginner's programming
\cite{hershkovitz2019creativity,israel2021associations}.

A challenge of classic creativity tests is that we need human experts
to count the number of generated ideas and decide which ones belong
to the same class. This reliance on human labor becomes infeasible
in scenarios where a large number of people engages in
creative activity, such as large-scale open learning environments
like Scratch \cite{Maloney2010}. The problem is compounded by the fact
that a single creativity measurement is likely not enough, given that
creativity is context-dependent and, thus, changes over time \cite{Amabile2018}.
Accordingly, our mission in this paper is to adapt the
definition of fluency, flexibility, and originality in two respects:
First, by making it applicable to student submissions in a 
multimodal learning environment, such as Scratch. Second, by making
it efficiently computable, without the need for human intervention.

There exists some prior work on
automating Torrance's test for online learning. In particular,
\cite{Huang2010} compute fluency, flexibility, and originality in
a collaborative brainstorming task; \cite{Yeh2015} automatically 
count the number of unique ideas in reaction to an 
inkblot-like picture; and \cite{Kovalkov2020,Kovalkov2021} automatically grade
the creativity of Scratch projects with a manually defined measure.
Our present work is a generalization of these prior approaches by
formalizing fluency, flexibility, and originality abstractly, based on
concepts and distances between them. This abstraction has the advantage
that we can transfer conceptual and computational approaches between
modalities or even domains with little need for adaptation.

The reason we use distances is twofold. First, distances are a
common representation in data mining and there is a rich toolbox of
distance measures which we can apply for our purposes \cite{Pekalska2005}.
More importantly, distances are helpful to model the organization
of knowledge in the mind. In particular, we can say that two
concepts have a low distance to each other if humans tend to
associate them more easily \cite{Hodgetts2009,Kenett2019}.
Two popular frameworks in this regard are \emph{semantic embeddings} and
\emph{semantic networks}. Semantic embeddings assume that concepts are implicitly represented
in a vector space and that their distance in this space corresponds to their semantic
relatedness \cite{Kenett2019,Landauer1998}.
By contrast, semantic networks assume that concepts are organized as a graph and that the shortest
path distance in the graph represents semantic relatedness
\cite{Kenett2019,Boden2004,Georgiev2018,Sowa2006}.
\cite{Kenett2019} suggests that both frameworks play a role in quantifying creativity
by providing complementary notions of distance between concepts.
We show later that our formalization is general enough to accommodate both frameworks.

We test our formalization in the example domain of Scratch. Scratch is an interactive,
block-based, and graphics-focused programming environment used for introductory
programming \cite{Maloney2010}.
Scratch is particularly interesting for us because Scratch projects
are intended for creative expression \cite{Bustillo2016,Giannakos2013}
and involve components from three different modalities,
namely code, visuals, and audio. Our ambition is to develop an 
automatic measure of creativity that can be applied to all three
modalities, whereas prior work has either disregarded image and
sound entirely \cite{hershkovitz2019creativity} or used only an
ad-hoc definition without theoretical justification
\cite{Kovalkov2020}.


\section{Formal Creativity Measure}
\label{sec:definition}

\tikzstyle{circ}=[draw=aluminium6, circle, inner sep=0.12cm, semithick]
\tikzstyle{sq}=[draw=aluminium6, rectangle, inner sep=0.15cm, semithick]
\tikzstyle{tri}=[draw=aluminium6, regular polygon, regular polygon sides=3, inner sep=0.07cm, yshift=-0.05cm,semithick]

\begin{figure*}
\begin{center}
\begin{tikzpicture}[black]

\begin{scope}[shift={(-3,0)}]
\node at (-0.4,0.8) {$\strut$ (a) Concept space};

\node[left] at (-0.7,0) {$\conspace = \{$};
\node[circ] at (-0.5,0) {};
\node at (-0.25,-0.1) {$,$};
\node[sq] at (0,0) {};
\node at (+0.25,-0.1) {$,$};
\node[tri] at (+0.5,0) {};
\node[right] at (+0.7,0) {$\}$};
\end{scope}

\begin{scope}[shift={(+1,-0.7)}]

\node at (0,1.5) {$\strut$ (b) Student submission $s$};

\draw[semithick, draw=aluminium6] (0,0.7) circle (0.2cm);
\draw[semithick, draw=aluminium6] (-0.25,0) rectangle (0.25,0.5);
\draw[semithick, draw=skyblue3] (-0.5,0.5) -- (-0.25,0.5) -- (-0.25,+0.25) -- cycle;
\draw[semithick, draw=scarletred3] (+0.5,0.5) -- (+0.25,0.5) -- (+0.25,+0.25) -- cycle;
\draw[semithick, draw=skyblue3] (-0.25,0) -- (0,0) -- (-0.25,-0.25) -- cycle;
\draw[semithick, draw=scarletred3] (+0.25,0) -- (0,0) -- (+0.25,-0.25) -- cycle;
\end{scope}

\begin{scope}[shift={(+7,-0.7)}]

\node at (0,1.5) {$\strut$ (c) reference submission $s'$};

\draw[semithick,draw=aluminium6] (-0.35,+0.35) -- (0,0.8) -- (+0.35,+0.35) -- cycle;
\draw[semithick,draw=aluminium6] (-0.35,-0.35) rectangle (0.35,0.35);
\end{scope}

\begin{scope}[shift={(-3.5,-3)}]

\node at (0,1.5) {$\strut$ (d) semantic network};

\node[tri] (tri)   at (-1,+0.5) {};
\node[sq]   (sq)   at (-1,-0.5) {};
\node     (zero)   at (0,0)    {0};
\node[circ] (circ) at (+1,0)    {};

\path[semithick, draw=aluminium6, shorten >=0.2cm, shorten <=0.2cm]
(tri) edge (sq) edge (zero)
(sq) edge (zero)
(zero) edge (circ);
\end{scope}

\begin{scope}[shift={(0.5,-3)}]

\node at (0,1.5) {$\strut$ (e) distance};

\node at (0,0) {
\begin{tabular}{ccccc}
$\delta(x,y)$ & \tikz\node[tri] {}; & \tikz\node[sq] {}; & \tikz\node[circ] {}; & 0 \\
\cmidrule(lr) {1-1} \cmidrule(lr) {2-4} \cmidrule(lr) {5-5}
\tikz\node[tri] {};  & 0 & 1 & 2 & 1 \\
\tikz\node[sq] {};   & 1 & 0 & 2 & 1 \\
\tikz\node[circ] {}; & 2 & 2 & 0 & 1 \\
0 & 1 & 1 & 1 & 0
\end{tabular}
};

\begin{scope}[shift={(2.5,0)}]

\node at (4,1.5) {$\strut$ (f) creativity measures};

\node[right] at (0,.8) {$\text{Flue}(s) = 4\cdot\delta(\tikz\node[tri]{};,0) + \delta(\tikz\node[sq]{};,0) + \delta(\tikz\node[circ]{};,0) = 6$};
\node[right] at (0,0) {$\text{Flex}(s) = \frac{1}{5} \Big[8\cdot\big(\delta(\tikz\node[tri]{};,\tikz\node[sq]{};) + \delta(\tikz\node[tri]{};,\tikz\node[circ]{};)\big) + 2 \cdot \delta(\tikz\node[sq]{};,\tikz\node[circ]{};)\Big] = \frac{28}{5}$};
\node[right] at (0,-.8) {$\text{Orig}(s) = 3\cdot\delta(\tikz\node[tri]{};,0) + \delta(\tikz\node[circ]{};,0) = 4$};
\end{scope}

\end{scope}
\end{tikzpicture}
\end{center}
\caption{Example creativity task illustrating our proposed
measure of creativity. (a) Participants receive a set of
three geometric shapes (triangle, rectangle, and circle) as
stimulus and have the task to form objects from these shapes.
(b) Example answer (a figure shaped out of a rectangle,
four triangles, and a circle; color is used to disambiguate triangles).
(c) Less creative answer (a house out of a rectangle and a triangle) as reference for originality.
(d) Possible semantic network including the triangle, square, and circle shape.
(e) Possible definition of $\delta$ for this task based on the
path distance in a semantic network.
(f) Fluency, flexibility, and originality of the figure (b)
according to $\delta$.}
\label{fig:creativity_example}
\end{figure*}
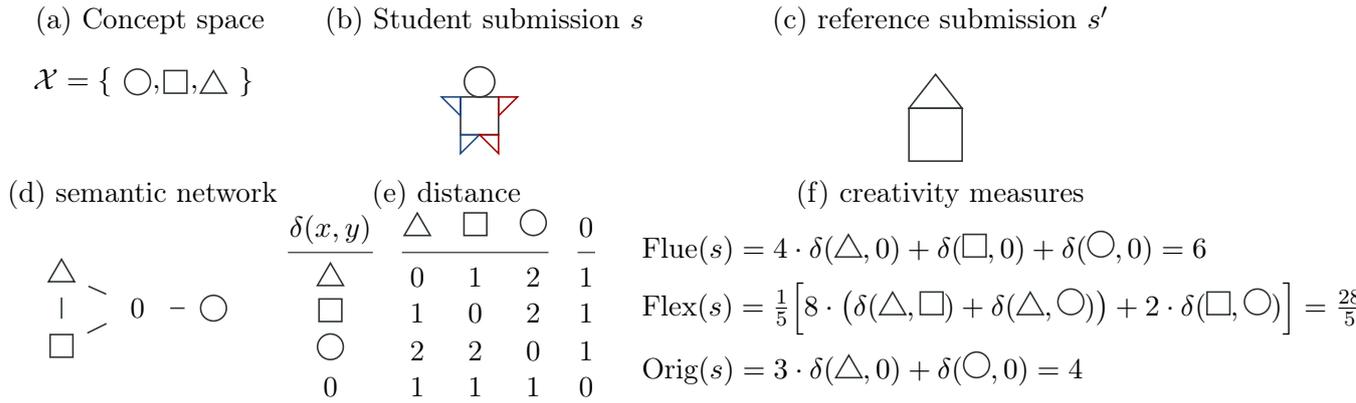

In this section, we introduce our proposed measure of creativity. Our main inspiration
is Torrance's test of creative thinking, which quantifies creativity by counting the
number of ideas in a creative product (fluency), the number of unique classes of ideas
(flexibility), and the infrequency of ideas compared to a reference population (originality)
\cite{Torrance1972,Kim2006}. We use these three scales since prior work has already
shown that they lend themselves for automation in learning environments
\cite{Huang2010,Kovalkov2020,Yeh2015} and are connected to success in
beginner's programming \cite{hershkovitz2019creativity,israel2021associations}. 
However, there are domains where a simple counting scheme may be insufficient.
Consider the example of Scratch projects (Fig.~\ref{fig:scratch}).
A Scratch project consists of code, images, and sounds.
If we merely count the number of code blocks, images, and sounds, we would loose 
a lot of information that could give us additional insight into the creativity of 
the project, for example, whether the code used advanced programming concepts or repeated
many simple operations. Instead, we propose to measure the difference between programming
concepts (and concepts more generally) by a distance metric. This distance, then, is
the basis for our formalizations of fluency, flexibility, and originality.

In general, our measure of creativity (Section~\ref{ssec:computing_creativity}) relies
on two ingredients, which we describe next: The concept space (Section~\ref{ssec:concept_space})
in which creative products live, and a distance between concepts (Section~\ref{ssec:distances}).

\subsection{Concept Spaces and Creative Products}
\label{ssec:concept_space}

We define a creative task as providing a student with a set of
building blocks, which the student can combine to a creative product.
We call these building blocks \emph{concepts} and the set of all available
concepts the \emph{concept space} $\conspace$. Consider the example in
Fig.~\ref{fig:creativity_example}. Here, the creative task is to combine
the three shapes circle, square, and triangle in a meaningful way.
Accordingly, the concept space is $\conspace = \{\tikz\node[circ]{};,\tikz\node[sq]{};,\tikz\node[tri]{};\}$.

Next, we define a \emph{creative product} $s$ as a combination of concepts.
More precisely, we say that $s$ is a graph
$s = (V_s, E_s)$, where $V_s \subseteq \conspace$ are the
concepts in $s$ and where $E_s \subseteq V_s \times V_s$
are the (syntactic) connections between concepts in $s$.
For example, the house in Fig.~\ref{fig:creativity_example}(c)
would be represented as the graph
$s' = \big(\{\tikz\node[sq]{};,\tikz\node[tri]{};\},
\{(\tikz\node[sq]{};, \tikz\node[tri]{};)\}\big)$,
because it involves a square and a triangle, where the square is
connected to the triangle. Similarly, the figure in Fig.~\ref{fig:creativity_example}(b)
would be represented as the graph
$s = \big(\{
\tikz\node[circ]{};,
\tikz\node[sq]{};,
\tikz\draw[semithick,skyblue3] (0,0) -- (0,0.3) -- (0.3,0.3) -- cycle;,
\tikz\draw[semithick,skyblue3] (0,0.3) -- (0.3,0.3) -- (0.3,0) -- cycle;,
\tikz\draw[semithick,scarletred3] (0,0) -- (0,0.3) -- (0.3,0.3) -- cycle;,
\tikz\draw[semithick,scarletred3] (0,0.3) -- (0.3,0.3) -- (0.3,0) -- cycle;
\}, \{
(\tikz\node[circ]{};, \tikz\node[sq]{};),
(\tikz\draw[semithick,skyblue3] (0,0) -- (0,0.3) -- (0.3,0.3) -- cycle;, \tikz\node[sq]{};),
(\tikz\draw[semithick,skyblue3] (0,0.3) -- (0.3,0.3) -- (0.3,0) -- cycle;, \tikz\node[sq]{};),
(\tikz\draw[semithick,scarletred3] (0,0) -- (0,0.3) -- (0.3,0.3) -- cycle;, \tikz\node[sq]{};),$
$(\tikz\draw[semithick,scarletred3] (0,0.3) -- (0.3,0.3) -- (0.3,0) -- cycle;, \tikz\node[sq]{};)
\}\big)$.

Prior literature has shown that a graph representation is flexible enough
to represent student work in a wide variety of domains and modalities \cite{Mokbel2013}.
For example, we can represent computer programs by syntax trees,
where the structure of the code is represented by the connections
within the trees \cite{Paassen2018JEDM,Price2017,Rivers2015};
we can represent images as a collection of depicted objects and 
their spatial relation via connections \cite{Johnson2015};
and we can represent audio data as a sequence of sounds, where
connections represent the temporal ordering.

Importantly, to automate creativity computation, we require an automatic way to
convert raw student output into a graph of concepts. Fortunately, such a conversion
is natural for many domains \cite{Mokbel2013}. For example, we describe the
conversion for Scratch projects in Section~\ref{sec:scratch_preprocess}.

\subsection{Distances}
\label{ssec:distances}

As a next step, we formalize the semantic relatedness between concepts
in a domain by means of a distance function $\delta$.
More precisely, we define a (semantic) distance $\delta$ as a function that
maps two concepts $x \in \conspace$ and $y \in \conspace$
to a real number, such that $\delta(x, y) \geq 0$,
$\delta(x, y) = \delta(y, x)$, and $\delta(x, y) = 0$ if and only if $x = y$.
These requirements stem from the mathematical definition of a distance metric
\cite{Pekalska2005}.
We use distances as an interface because they are sufficient to define
creativity but abstract enough to be easily adaptable across domains and modalities
\cite{Pekalska2005}. Additionally, distances connect nicely to prior
research in cognitive science. To illustrate this connection, we provide
two examples of distances.

First, we consider semantic networks \cite{Boden2004,Sowa2006}.
The theory of semantic networks suggests that human cognition arranges
concepts in a graph, where connections express semantic relatedness.
We can obtain a distance $\delta$ from a semantic network by
setting the concept space $\conspace$ to the nodes of the network and
$\delta$ to the minimum number of edges we need to traverse to get from
one concept to another \cite{Georgiev2018}.
Fig.~\ref{fig:creativity_example}(d) shows a very
simple example of such a semantic network. In this network, we connect
triangle and square because they are related---for example, we can construct a
square out of two triangles and the square is the next regular polygon
by number of vertices. By contrast, the circle is not directly related
to either shape, which is why we do not connect it directly but only
via a 'neutral shape' $0$. Accordingly, the distance between triangle
and square is $\delta(\tikz\node[tri]{};,\tikz\node[sq]{};) = 1$, because
these shapes are directly connected in the network, and
$\delta(\tikz\node[tri]{};,\tikz\node[circ]{};) = 2$ because we
need to make two hops to get from the triangle to the circle.
The matrix of all pairwise distances is shown in Fig.~\ref{fig:creativity_example}(e).
In Fig.~\ref{fig:scratch_blocks}, we use a semantic network to represent the relatedness between code
blocks in Scratch.

Second, we consider semantic embeddings \cite{Kenett2019,Landauer1998}.
Semantic embeddings assume that concepts have an implicit representation in a high-dimensional
vector space and that their semantic relatedness corresponds to the Euclidean distance
or the cosine similarity between the vectors. Accordingly, we can set $\delta$
to the Euclidean distance $\delta(\vec x, \vec y) = \lVert \vec x - \vec y\rVert_2$
or the cosine distance $\delta(\vec x, \vec y) =
1 - \frac{\vec x^T \cdot \vec y}{\lVert \vec x \rVert \cdot \lVert \vec y \rVert}$.
In Section~\ref{sec:scratch_preprocess}, we use the Euclidean distance for sounds and
the cosine distance for images in Scratch.

A final feature of our proposed distance-based framework is that we can express
the amount of domain knowledge expressed by a concept via its distance $\delta(x, 0)$ to a
neutral nullconcept $0$. For example, in Scratch code we use the
distance $\delta(x, 0)$ to express how advanced a programming concept is
(Section~\ref{sec:scratch_preprocess}).

\subsection{Computing Creativity}
\label{ssec:computing_creativity}

In this section, we adapt Torrance's scales of fluency, flexibility,
and originality \cite{Torrance1972} to our formalization.
First, let us recall how Torrance's test works: We present participants with
a prompt in form of a concept space $\mathcal{X}$ and then ask them to generate as many creative
products over this concept space as possible. The number of generated creative products is called fluency,
the number of distinct classes of generated products is called flexibility, and the infrequency
of products in comparison to a general population is called originality.

Our aim here is slightly different: We consider a single creative product and wish to quantify the
amount of creativity expressed by this product in terms of fluency, flexibility, and originality.
Fortunately, the three scales still apply if we count concepts instead of products. In more
detail, we obtain the following three scales.

\subsubsection{Fluency} Torrance's test defines fluency as the number of generated ideas \cite{Kim2006,Torrance1972}.
Applied to a single creative product, this would mean that a product expresses more fluency
if it contains more concepts.
Thanks to our distance-based framework, we can be more nuanced. We formalize the
\emph{amount} of fluency expressed by a concept $x$ by its distance $\delta(x, 0)$ to the
nullconcept. The fluency of a product is the sum over all these distances within the product:
\begin{equation}
	\mathrm{Flue}(s) = \sum_{x \in V_s} \delta(x, 0) \label{eq:flu}
\end{equation}
Note that this is a proper generalization over counting the number of concepts
because we can set $\delta(x, 0) = 1$ for all concepts $x$, yielding $\mathrm{Flue}(s) = |V_s|$.
Computationally, fluency is efficient because it only requires 
$|V_s|$ distance computations.

For example, the fluency of the figure in Fig.~\ref{fig:creativity_example}(b) is $6$
because it consists of six shapes, each of which has distance $1$ to the nullconcept (Fig.~\ref{fig:creativity_example}(f)).

\subsubsection{Flexibility} Torrance's test defines flexibility as the 
number of distinct classes of ideas that are generated 
\cite{Kim2006,Torrance1972}. In other words, if a participant only generates ideas
that belong to the same class---for example, different houses for the drawing task in
Fig.~\ref{fig:creativity_example}---then this would be counted as a flexibility of
one. Each additional distinct class of ideas increases flexibility by one.
According to \cite{Kim2006}, this definition is inspired by Guilford's structure of
intellect model \cite{Guilford1956} and is meant to capture the diversity of generated ideas.
Fortunately, our framework enables us to capture the diversity of concepts
in more nuance than a binary choice whether it's distinct or not. In particular, we can
set the distance $\delta(x, y)$ such that it quantifies \emph{how} distinct $x$ is from $y$.
Once we have set up $\delta$ in this way, we formalize flexibility as the sum of all
pairwise distances between concepts, normalized by a factor of $|V_s|-1$ to 
maintain the same scale as for fluency.
\begin{equation}
	\mathrm{Flex}(s) = \frac{1}{|V_s|-1} \sum_{x \in V_s} \sum_{y \in V_s} \delta(x, y) \label{eq:flex}
\end{equation}
For flexibility, it makes sense to preprocess $V_s$ and remove duplicates. This conforms to Torrance's notion that flexibility should represent distinct classes of ideas. With this preprocessing step, flexibility is a proper generalization over counting the number of distinct concepts because we can set $\delta(x, y) = 1$ if $x \neq y$ and
$\delta(x, x) = 0$, yielding a flexibility of $\mathrm{Flex}(s) = |\tilde V_s|$, where $\tilde V_s$ refers to the duplicate-free version of $V_s$.
Regarding computational complexity, flexibility requires a sum of
pairwise distances, which is in $\mathcal{O}(|V_s|^2)$.

For example, the figure in Fig.~\ref{fig:creativity_example}(b) yields a
flexibility of $\frac{28}{5}$, because it consists
of four triangles, one square, and one circle, and we hence add
four times the distance $\delta(\tikz\node[tri]{};, \tikz\node[sq]{};) = 1$,
four times $\delta(\tikz\node[sq]{};, \tikz\node[tri]{};) = 1$,
four times $\delta(\tikz\node[tri]{};, \tikz\node[circ]{};) = 2$,
four times $\delta(\tikz\node[circ]{};, \tikz\node[tri]{};) = 2$,
one time $\delta(\tikz\node[sq]{};, \tikz\node[circ]{};) = 2$,
and one time $\delta(\tikz\node[circ]{};, \tikz\node[sq]{};) = 2$,
yielding $8 \cdot 1 + 10 \cdot 2 = 28$. Then, we normalize by the
number of shapes minus one, yielding $\frac{28}{5}$ (Fig.~\ref{fig:creativity_example}(f)).

\subsubsection{Originality} Torrance's test defines originality as the statistical infrequency of
ideas in comparison to a general population \cite{Kim2006,Torrance1972}. In other words,
one first needs to calibrate the test using a sample of participants from the general population,
yielding a frequency $f(x)$ for each idea $x$. Then, for
each new participant, we can determine the \emph{infrequency} of each idea as $1 - f(x)$
and add up these infrequencies to obtain originality.

A shortcoming of this definition is that it can not distinguish the originality of
two ideas which never occurred before. Consider our example in Fig.~\ref{fig:creativity_example}
and let's compare the figure in Fig.~\ref{fig:creativity_example}(b) with a variation of the house
in Fig.~\ref{fig:creativity_example}(b), where we just add
another square to represent the door of the house. Both shapes, the figure and the
altered house, are distinct from the house $s'$---but the figure is arguably \emph{more}
original because it is less similar to the house.
This is in line with \cite{Runco2012} who notes that distance to previously existing ideas
can be seen as particular evidence of genius in creative achievement.

Overall, we define the originality of a creative product $s$ as the average distance 
$d_\delta(s, s')$ to typical products $s'$ in a sample $S$.
\begin{equation}
\mathrm{Orig}(s) = \frac{1}{|S|} \sum_{s' \in S} d_\delta(s, s'). \label{eq:orig}
\end{equation}
Note that, as in Torrance's definition, we require a sample of typical products $S$
to define originality. Indeed, this appears to be a fundamental property of 
originality: We always need a reference point with respect to which
we define something as original \cite{Runco2012}. For example, \cite{Boden2004} distinguishes between
creativity with respect to my own prior ideas and creativity with respect to the entire prior
human history. In education, \cite{Spendlove2008} suggests to focus on \enquote{ordinary} creativity 
which can regularly be observed, rather than exceptional genius.
Accordingly, the sample $S$ should be chosen to represent products one would expect
of a typical student in the same context. For example, it could be a sample of products from students
of a similar age, school system, and socio-economic background.
Importantly, we may need different samples to account for different contexts appropriately
\cite{Amabile2018}.

Another important point is that originality requires a distance $d_\delta$ between products,
which we define next.

\subsubsection{Distances between products} Given a distance $\delta$ between concepts,
how can we compute a distance $d_\delta$ between two creative products $s$ and $s'$?
Here, we take inspiration from cognitive science.
\cite{Hodgetts2009} reviewed how humans perceive similarity between objects that consist
of parts and found two major notions in the literature: Structural alignment and
representational distortion.

Structural alignment \cite{Markman1993} measures distance by the difficulty of aligning
the parts of both objects. For example, consider the figure $s$ in Fig.~\ref{fig:creativity_example}(b)
and the house $s'$ in Fig.~\ref{fig:creativity_example}(c). We can align the square and one triangle
of both objects, but the remaining three triangles and the circle of the figure
can not be aligned, meaning that structural alignment would tell us that $s$ and $s'$
are fairly dissimilar. However, when we compare the figure $s$ to itself, we can
align every shape, meaning that the distance is zero.

Representational distortion \cite{Chater1997} measures distance by the effort needed
to transform one object into another by a sequence of predefined transformations.
For example, we can transform the figure $s$ into the house $s'$ by removing the
circle and three triangles and putting the remaining triangle on top of the square.

In general, both structural alignment and representational distortion are hard
to compute.
Structural alignment requires a search over the combinatorially large space of possible
alignments between two objects to find the best one, and representational distortion
requires a search over the infinitely large space of possible transformation sequences
that turn one object into another. Fortunately, algorithmic research in the past decades
has found special cases where these searches become efficient, namely \emph{edit distances} 
\cite{Bille2005,Bougleux2017,Levenshtein1965,Paassen2018JEDM,Zhang1989}.

To define an edit distance, we start from structural alignment.
In particular, we define an \emph{alignment} between two products $s$ and $s'$ as a
set of tuples $M \subseteq (V_s \cup \{0\}) \times (V_{s'} \cup \{0\})$ such that
any concept in $V_s$ occurs exactly once on the left-hand-side and any concept in
$V_{s'}$ occurs exactly once on the right-hand-side.
Consider again the example of the figure $s$ in Fig.~\ref{fig:creativity_example}(b)
and the house $s'$ in Fig.~\ref{fig:creativity_example}(c). One possible alignment between
$s$ and $s'$ would be $M = \{
(\tikz\draw[semithick,skyblue3] (0,0) -- (0,0.3) -- (0.3,0.3) -- cycle;, \tikz\node[tri]{};),
(\tikz\draw[semithick,skyblue3] (0,0.3) -- (0.3,0.3) -- (0.3,0) -- cycle;, 0),
(\tikz\draw[semithick,scarletred3] (0,0) -- (0,0.3) -- (0.3,0.3) -- cycle;, 0),
(\tikz\draw[semithick,scarletred3] (0,0.3) -- (0.3,0.3) -- (0.3,0) -- cycle;, 0),
(\tikz\node[sq]{};, \tikz\node[sq]{};),
(\tikz\node[circ]{};, 0)
\}$, that is, we align one of the four limbs of the figure
\tikz\draw[semithick,skyblue3] (0,0) -- (0,0.3) -- (0.3,0.3) -- cycle;
to the roof of the house \tikz\node[tri]{}; and the torso of the figure \tikz\node[sq]{};
to the house's wall \tikz\node[sq]{};. All remaining shapes in $s$, namely
\tikz\draw[semithick,skyblue3] (0,0.3) -- (0.3,0.3) -- (0.3,0) -- cycle;,
\tikz\draw[semithick,scarletred3] (0,0) -- (0,0.3) -- (0.3,0.3) -- cycle;, 
\tikz\draw[semithick,scarletred3] (0,0.3) -- (0.3,0.3) -- (0.3,0) -- cycle;, and
\tikz\node[circ]{};,
can not be aligned to anything in $s'$ anymore, such that we must align them to the nullconcept instead.
Note that other alignments are possible as well, for example
$M' = \{
(\tikz\draw[semithick,skyblue3] (0,0) -- (0,0.3) -- (0.3,0.3) -- cycle;, \tikz\node[sq]{};),
(\tikz\draw[semithick,skyblue3] (0,0.3) -- (0.3,0.3) -- (0.3,0) -- cycle;, 0),
(\tikz\draw[semithick,scarletred3] (0,0) -- (0,0.3) -- (0.3,0.3) -- cycle;, 0),
(\tikz\draw[semithick,scarletred3] (0,0.3) -- (0.3,0.3) -- (0.3,0) -- cycle;, 0),
(\tikz\node[sq]{};, 0),
(\tikz\node[circ]{};, \tikz\node[tri]{};)
\}$, where one of the limbs of the figure is aligned with the wall of the house and the
head of the figure is aligned with the roof of the house. Intuitively, the latter alignment is
worse because we do not align similar concepts of both products, even though we could.
Edit distances follow the same intuition: We define the cost $c_\delta(M)$ of an alignment $M$ as the sum
over the distances between aligned parts and we define the edit distance $d_\delta(s, s')$
as the cost of the cheapest alignment between $s$ and $s'$. More formally, we obtain:
\begin{equation}
d_\delta(s, s') = \text{\hspace{-0.3cm}}\min_{M \in \mathcal{M}(s, s')} \text{\hspace{-0.3cm}}c_\delta(M),
\text{ where } c_\delta(M) = \text{\hspace{-0.3cm}}\sum_{(x, y) \in M} \text{\hspace{-0.3cm}}\delta(x, y), \label{eq:match_dist}
\end{equation}
and where $\mathcal{M}(s, s')$ denotes the set of all alignments between $s$ and $s'$.

Returning to our example above, we see that $M$ has the cost
$c_\delta(M) = 3 \cdot \delta(\tikz\node[tri]{};, 0) + \delta(\tikz\node[circ]{};, 0) = 4$
and $M'$ has the cost $c_\delta(M') = \delta(\tikz\node[tri]{};, \tikz\node[sq]{};)
+ 3 \cdot \delta(\tikz\node[tri]{};, 0)
+ \delta(\tikz\node[sq]{};, 0)
+ \delta(\tikz\node[circ]{};, \tikz\node[tri]{};) = 7$, that is, $M$ is cheaper than $M'$.
Indeed, $M$ is the cheapest possible alignment between $s$ and $s'$, meaning
that $c_\delta(M) = d_\delta(s, s') = 3$.
Provided that $s'$ is the only product in our sample $S$, this value
also expresses the originality of $s$ (Fig.~\ref{fig:creativity_example}(f)).

It can be shown that the cheapest alignment $M$ can be found automatically in
$\mathcal{O}((|V_s| + |V_{s'}|)^3)$ using the Hungarian algorithm \cite{Munkres1957}.
Even better, it is possible to restrict the set of alignments $\mathcal{M}$ to take
specifics of our data into account.
For example, for sound we want to make sure that the alignment preserves the
temporal order, yielding the sequence edit distance of \cite{Levenshtein1965}
with the complexity $\mathcal{O}(|V_s| \cdot |V_{s'}|)$. For code, we want that the tree
structure of the program syntax is maintained, yielding the tree edit distance of \cite{Zhang1989}
with complexity $\mathcal{O}(|V_s|^2 \cdot |V_{s'}|^2)$.

Beyond their computational appeal, edit distances have another advantage: they unify
structural alignment and representational distortion. This is because we can convert
any alignment between two products $s$ and $s'$ to a sequence of transformations from
$s$ to $s'$ and vice versa. In particular, any tuple $(x, 0) \in M$ corresponds to a
deletion of concept $x$ from $s$, any tuple $(0, y) \in M$ corresponds to an insertion
of concept $y$ into $s$, and any tuple $(x, y) \in M$ corresponds to a replacement of
concept $x$ with concept $y$. In this translation, costs are preserved, such that
the cheapest alignment also corresponds to the cheapest sequence of deletions, insertions,
and replacements between $s$ and $s'$ \cite{Bougleux2017,Zhang1989}.
Hence, structural alignment and representational distortion become equivalent.

We note that we are not the first to apply edit distances on educational data.
Edit distances are a typical tool to compare computer programs \cite{Paassen2018JEDM,Price2017,Rivers2015}
and have previously been suggested as a domain-independent measure of proximity
for educational data \cite{Mokbel2013}. To us, this is additional encouragement
that edit distances are a natural way to compute originality in educational settings.

Finally, we note that \ref{eq:orig} is a proper generalization of Torrance's
definition in terms of infrequency. In particular, we recover infrequency as a special
case by setting $d_\delta$ to the alignment distance in
\ref{eq:match_dist}, and by setting $\delta(x, x) = \delta(0, x) = 0$ with $\delta(x, y) = 1$,
otherwise. For a proof, refer to the supplementary material.

This concludes our formalization of creativity.
We will now turn to the domain of Scratch and elaborate on how we
implement our proposed creativity measure for the three modalities of code,
images, and sound.

\section{Creativity in Scratch}
\label{sec:scratch_preprocess}

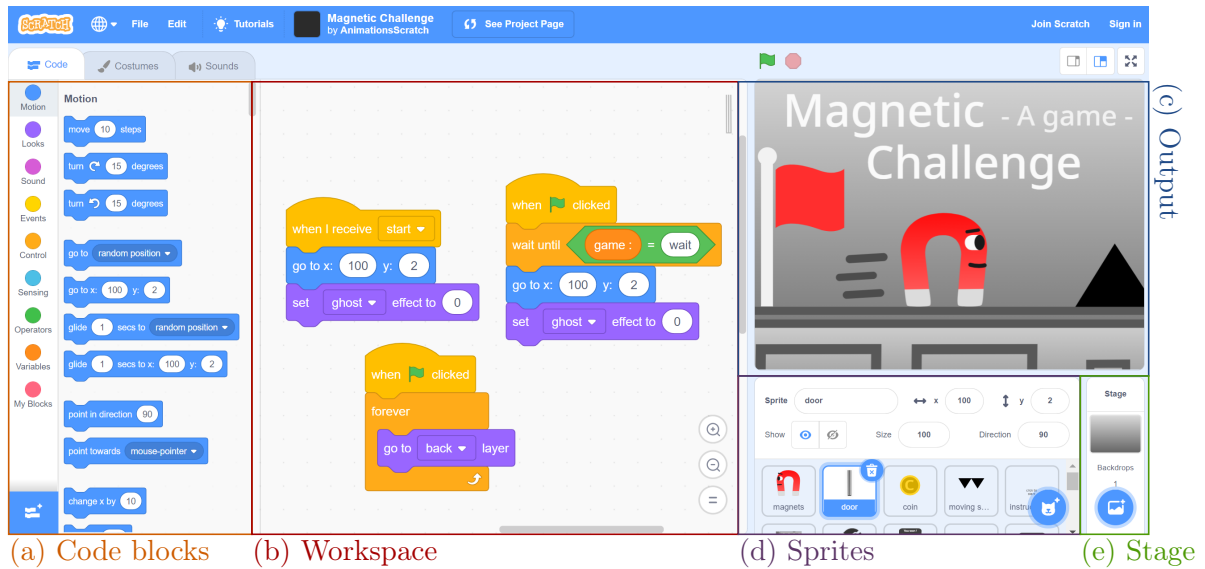
\begin{figure*}
\begin{center}
\begin{tikzpicture}
\node at (0,0) {\includegraphics[width=15.cm]{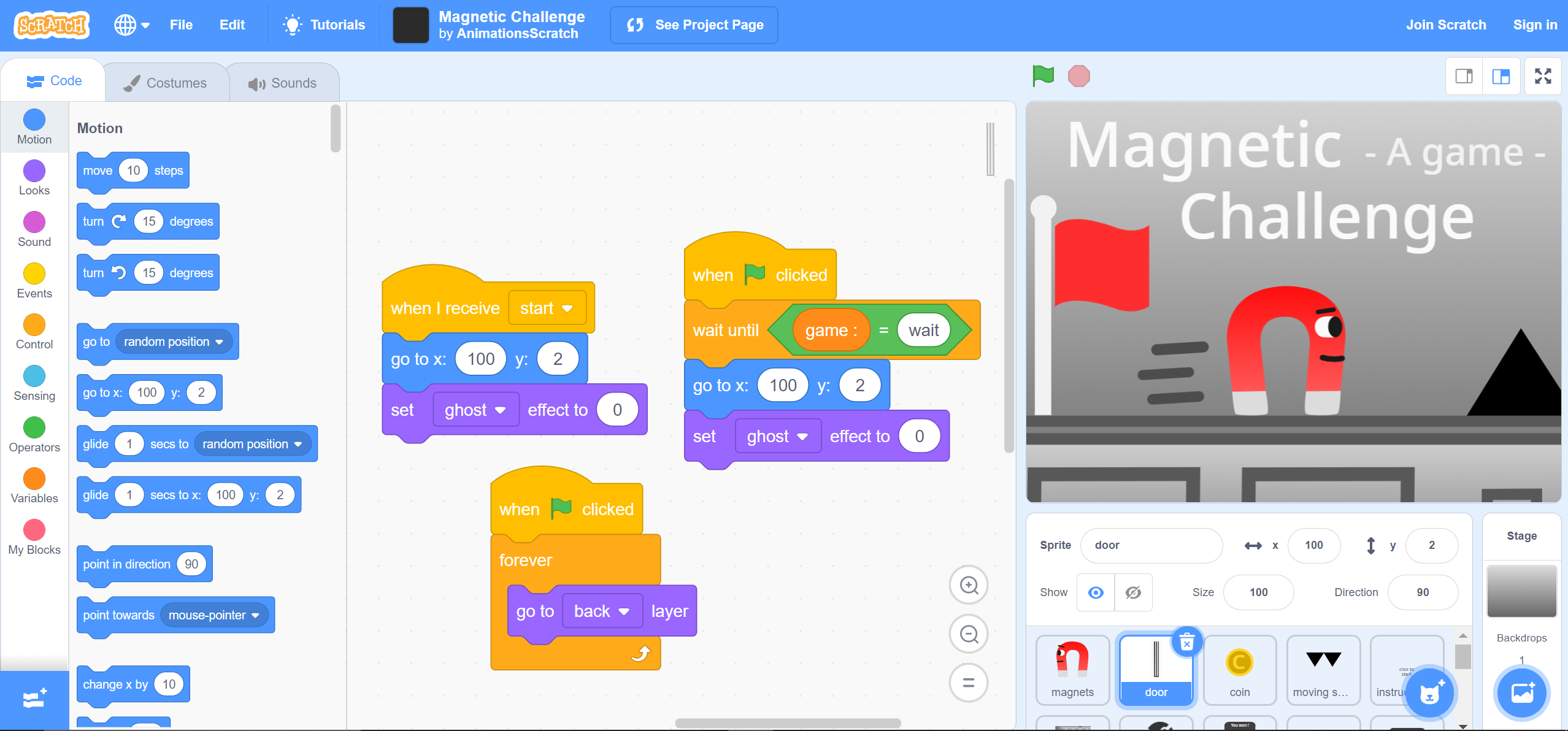}};

\draw[orange3, semithick] (-7.5,-3.5) rectangle (-4.3,+2.5);
\node[orange3, right, inner sep=0pt] at (-7.5,-3.75) {(a) Code blocks$\strut$};

\draw[scarletred3, semithick] (-4.3,-3.5) rectangle (+2.1,+2.5);
\node[scarletred3, right, inner sep=0pt] at (-4.3,-3.75) {(b) Workspace$\strut$};

\draw[skyblue3, semithick] (2.1,-1.4) rectangle (+7.5,+2.5);
\node[skyblue3, above right, rotate=270, inner sep=0pt] at (+7.5,+2.5) {(c) Output$\strut$};

\draw[plum3, semithick] (2.1,-3.5) rectangle (+6.6,-1.4);
\node[plum3, right, inner sep=0pt] at (2.1,-3.75) {(d) Sprites$\strut$};

\draw[chameleon3, semithick] (+6.6,-3.5) rectangle (+7.5,-1.4);
\node[chameleon3, right, inner sep=0pt] at (+6.6,-3.75) {(e) Stage$\strut$};

\end{tikzpicture}
\end{center}
\caption{Screenshot of an example Scratch project. (a) Overview of possible code blocks.
(b) Code blocks related to the currently selected sprite (the door). (c) Current output. (d) Overview of all sprites. (e) Stage/background.}
\label{fig:scratch}
\end{figure*}

Scratch is a block-based programming environment with a strong focus
on visual output that is used for introductory programming \cite{Maloney2010}.
Fig.~\ref{fig:scratch} presents a sample Scratch project, a game called \enquote{Magnetic Challenge.}
In this game, the player's task is to steer a magnet across a path of spike obstacles as
fast as possible while collecting coins, using the keyboard keys or the mouse.
As shown in Fig.~\ref{fig:scratch}, the Scratch interface is divided into five sections:
In the stage area (Fig.~\ref{fig:scratch}(e)), the user can select a graphical background for their
project. In the sprites area (Fig.~\ref{fig:scratch}(d)), the user can add new foreground images 
(sprites) to their project. Such sprites can be drawn manually, chosen from the Scratch library, or 
uploaded. The code blocks area (Fig.~\ref{fig:scratch}(a)) provides an overview of all possible code
that the user may attach to sprites or backgrounds. Blocks are grouped into predefined categories, such
as \enquote{motion,} \enquote{looks,} and \enquote{events.} Additionally, a user can use the extension category 
to load additional block packages with advanced functionality, such as \enquote{video} and
\enquote{text to speech.} Finally, users can create custom blocks, which will appear in \enquote{My Blocks.}

The workspace area (Fig.~\ref{fig:scratch}(b)) displays all code blocks associated with the currently
selected sprite or background. Users can drag and drop blocks from the code blocks area to the workspace
and combine them in the workspace as needed. Each block combination starts with an event and continues
with blocks that are executed sequentually whenever the event occurs. For example, 
Fig.~\ref{fig:scratch}(b) displays all blocks associated with the \enquote{door} sprite in the 
\enquote{Magnetic Challenge} game.
Finally, the output area  (Fig.~\ref{fig:scratch}(c)) executes the project such that the user can interact with it.

\definecolor{events}{HTML}{ffbf00}
\definecolor{events-outline}{HTML}{cc9900}
\definecolor{motion}{HTML}{4c97ff}
\definecolor{motion-outline}{HTML}{397bd8}
\definecolor{extension}{HTML}{0fbd8c}
\definecolor{extension-outline}{HTML}{0b8e69}
\definecolor{custom}{HTML}{ff6680}
\definecolor{custom-outline}{HTML}{ff405f}

\begin{figure}
\begin{center}
\begin{tikzpicture}
\node[circle, draw=aluminium6, fill=aluminium4] (null) at (0,0) {0};

\node[rectangle, rounded corners] (basic) at (0,-1.5) {predefined};

\node[right] (events) at (1.5,-0.5) {\includegraphics[scale=0.2]{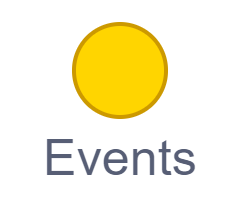}};
\node[right] (when_pressed) at (4,-0.5) {\includegraphics[scale=0.2]{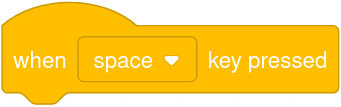}};
\node at (6,-1) {$\vdots$};

\node[right] (motion) at (1.5,-1.5) {\includegraphics[scale=0.2]{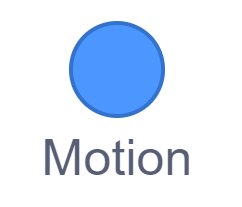}};
\node[right] (move) at (4,-1.5) {\includegraphics[scale=0.2]{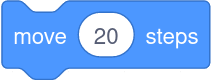}};
\node at (5,-2) {$\vdots$};

\node at (2,-2) {$\vdots$};

\node[rectangle, rounded corners] (extensions) at (0,-3) {extensions};

\node[right] (pen) at (1.4,-3) {\includegraphics[scale=0.3]{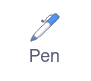}};
\node[right] (pen_down) at (4,-3) {\includegraphics[scale=0.2]{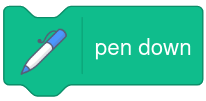}};
\node at (5,-3.5) {$\vdots$};

\node at (2,-3.5) {$\vdots$};

\node[rectangle, rounded corners, fill=custom, draw=custom-outline, text=white] (custom) at (0,-4.5) {custom blocks};

\node[right] (myblock) at (2.5,-4.5) {\includegraphics[scale=0.2]{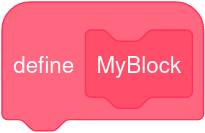}};

\path[-, semithick, shorten >=0.1cm, shorten <=0.1cm]
(null) edge node[left] {2} (basic)
(basic) edge node[left] {1} (extensions) edge node[above left] {$0.5$} (events) edge node[above] {$0.5$} (motion)
(events) edge[draw=events-outline] node[above] {$0.5$} (when_pressed)
(motion) edge[draw=motion-outline] node[above] {$0.5$} (move)
(extensions) edge node[left] {1} (custom) edge node[above] {$0.5$} (pen)
(pen) edge[draw=extension-outline] node[above] {$0.5$} (pen_down)
(custom) edge[draw=custom-outline] node[above] {$1$} (myblock);
\end{tikzpicture}
\end{center}
\caption{Semantic network, organizing the concept space $\mathcal{X}$ and distance $\delta$ for Scratch blocks.
Code blocks are distinguished into predefined, extension, and custom blocks, with 
subcategories for predefined (events, motion, $\ldots$) and
extensions (pen, translate, $\ldots$). The distance between blocks corresponds 
to the shortest path distance in the network. For example,
$\delta(\includegraphics[scale=0.1]{scratch_buttons/move.png}, \includegraphics[scale=0.1]{scratch_buttons/pen_down.png})
= 0.5 + 0.5 + 1 + 0.5 + 0.5 = 3$.}
\label{fig:scratch_blocks}
\end{figure}

Importantly, Scratch programs are multimodal, including code, images, and sound. To fully account
for the creativity in a Scratch project, we want to include all three modalities in our creativity
measure. We do so by applying our framework from Section~\ref{sec:definition} to each modality separately, meaning that for code, images, and sound we obtain a measure of fluency, flexibility, and originality, repectively. The fact that our framework applies to all three modalities attests to the generality of our methods. In Section~\ref{sec:predict}, we combine fluency, flexibility, and originality from code, images, and sound into an overall creativity measure via a machine learning model. This model is trained to match human creativity assessments.

\subsubsection{Code creativity}
Similar to previous works, we represent a Scratch project's code as a collection of syntax trees \cite{Price2017} one
representing the stage, and one per sprite. The concept space $\conspace$ contains all
predefined blocks and extension blocks in the Scratch language, as well as user-defined custom blocks, 
and the null concept. We define the distance metric $\delta$ as the shortest-path distance in the semantic network in Fig.~\ref{fig:scratch_blocks}. Our network models the hierarchy of Scratch blocks, where we distinguish between predefined blocks, extension blocks, and custom blocks. Within each category, we further distinguish subcategories as defined by the Scratch language. For predefined blocks, these are motion, events, control, etc. (Fig.~\ref{fig:scratch}(a)). For extensions, these are the single extension packages, such as video and text to speech. For custom blocks, the Scratch environment does not provide subcategories.

The edge labels in Fig.~\ref{fig:scratch_blocks} represent the length of each edge.
The distance between any two blocks equals sum over all edges that we need to traverse
to get from one block to the other. For example, the distance between the \enquote{move} block and
the \enquote{when key pressed} block is two due to the following calculation.
\begin{align*}
\delta(\includegraphics[scale=0.05]{scratch_buttons/move.png}, \includegraphics[scale=0.05]{scratch_buttons/when_pressed.png}) = &\delta(\includegraphics[scale=0.05]{scratch_buttons/move.png}, \includegraphics[scale=0.15]{scratch_buttons/motion.PNG}) +
\delta(\includegraphics[scale=0.15]{scratch_buttons/motion.PNG}, \text{predefined}) + \\
&\delta(\text{predefined}, \includegraphics[scale=0.15]{scratch_buttons/events.PNG}) + \delta(\includegraphics[scale=0.15]{scratch_buttons/events.PNG}, \includegraphics[scale=0.05]{scratch_buttons/when_pressed.png}) = 0.5\cdot4
\end{align*}
Overall, we obtain a distance of one between blocks in the same subcategory, a
distance of two between blocks in different subcategories but the same category,
a distance of three between predefined and extension blocks or between extension blocks
and custom blocks, and a distance of four between predefined and custom blocks.

The distance to the nullconcept reflects how much programming knowledge and effort is required to
use the block. For predefined blocks, we obtain a distance of three, for extension blocks a distance of
four, and for custom blocks a distance of five.

Based on this distance, we computed code fluency, flexibility, and originality of $45$ Scratch projects. 
Table~\ref{tbl:code-creativity-stat} presents the resulting statistics.

We compute fluency via \ref{eq:flu}, which counts how many blocks
are contained in a program, with a custom block being \enquote{worth} five points of fluency,
an extension block four points, and a predefined block three points. For example, in the \enquote{Magnetic Challenge} program, the \enquote{go to} block presented in Fig.~\ref{fig:scratch} is a predefined block from class \enquote{motion}. The program gets three points for this block, and an overall fluency score of $21\,484$, which is higher then the average fluency score of the projects we assessed. The high number occurs because we use squared $\delta$ in practice, which has the added value of making fluency interpretable as a squared length, and flexibility as a variance.

For the flexibility measure, we first remove all duplicated blocks as recommended in Section~\ref{ssec:computing_creativity}. For example, the project \enquote{Magnetic Challenge} uses the \enquote{go to} block and \enquote{when flag clicked} block more than once (Fig.~\ref{fig:scratch}(b)).
Then, we compute \ref{eq:flex}. For example, the distance between the \enquote{go to} block and the
\enquote{when flag clicked} block contributes two to flexibility. Overall, after normalizing by the number of the unique blocks $|\tilde V_s| = 72$ in the program, it obtains a flexibility score of $477.92$ which is higher then the average flexibility score of the projects we assessed.

Originality requires a sample of projects, meaning we need a reference point with respect to which a project is original or not. This follows prior work by \cite{Torrance1972,Runco2012}.
Since the choice of sample can influence originality, we considered three reference samples, two with 20 programs, one with 10 programs. The samples were chosen randomly out of our 45 unique Scratch projects
with overlap. Depending on the sample, the \enquote{Magnetic Challenge} program in Fig.~\ref{fig:scratch} got three different scores for originality: $7023.05$, $6641.53$, and $3260.5$. The average originality across projects is $4105.02$ (see Table~\ref{tbl:code-creativity-stat}), indicating that the code originality of \enquote{Magnetic Challenge} is relatively high.

Further, to compute originality as defined in Section~\ref{ssec:computing_creativity}, we need an alignment distance between Scratch projects. 
To do so, we follow the three-step approach of \cite{Price2017} for block-based programming.
First, we compute the tree edit distance \cite{Zhang1989} between
the stage syntax trees of both programs. Then, we compute all pairwise tree edit distances
between sprites in both programs. Finally, we feed this result into the Hungarian
algorithm \cite{Munkres1957} to obtain an optimal matching between the sprites.
Regarding computational efficiency, we obtain a time complexity of
$\mathcal{O}(m_\text{stage}^2 \cdot n_\text{stage}^2 + m \cdot n \cdot m_\text{sprite}^2 \cdot n_\text{sprite}^2  + (m+n)^3)$
where $m_\text{stage}$ and $n_\text{stage}$ are the number of blocks in the stage,
$m_\text{sprite}$ and $n_\text{sprite}$ are the maximum number of blocks per sprite,
and $m$ and $n$ are the number of sprites. Because all these numbers are usually quite
small in Scratch projects, this computation is still fast \cite{Price2017}.

\begin{table}
\caption{Code Creativity Statistics}
\label{tbl:code-creativity-stat}
\centering
\begin{tabular}{cccc}
\toprule
\midrule
& fluency &  flexibility &  originality \\
\midrule
mean & $13\,342.46$ & $372.16$ & $4105.02$ \\
std  & $19\,398.40$ & $128.60$ & $2848.86$ \\
min  & $883$        & $89.75$  & $834$ \\
max  & $119\,849$   & $747.82$ & $23\,468$ \\
\midrule
\bottomrule
\end{tabular}
\end{table}

\subsubsection{Visual creativity}
Projects in Scratch may contain different images, either provided by Scratch, drawn by the users, or 
uploaded by them. The concept space $\mathcal{X}$ consists of all possible images that could be
included in a project. Note that this space is infinitely large and varied, including images of various 
sizes, file formats, content, etc. Accordingly, we decide to apply a semantic embedding approach, where
we embed images in a shared space before we compare them \cite{Kenett2019,Landauer1998}. But what is an
appropriate, semantic embedding for images? In computer vision, tremendous progress has been achieved
via deep neural networks \cite{feng2019computer}, which have achieved state-of-the-art performance
in challenging tasks such as object detection~\cite{lin2019transfer}, image classification~\cite{xia2017inception}, and fault diagnosis~\cite{wen2019new}. Note that all these tasks concern the semantics of an image, such that deep neural networks appear as a useful tool for our purpose. In particular, we select ResNet50 \cite{He2016}.
ResNet50 is a convolutional deep neural network with residual connections,
which achieved state-of-the-art results in the famous ImageNet competition \cite{He2016}.
It has been shown to capture semantic differences between images in tasks such as medical
image classification \cite{Reddy2019}, software classification \cite{Rezende2017}, and many more.

ResNet50 translates each image into a vector. Accordingly, we 
represent the images in a Scratch project as a set of vectors, one per image. 
To compare the image vectors, we chose the Cosine distance  $\delta(\vec x, \vec y) =
1 - \frac{\vec x^T \cdot \vec y}{\lVert \vec x \rVert \cdot \lVert \vec y \rVert}$ 
because it is invariant against effects of scale and size.
Using this distance, we computed the visual fluency, flexibility, and originality for 45 Scratch projects. Table~\ref{tbl:visual-creativity-stat} presents the resulting statistics.

Visual fluency is computed via \ref{eq:flu}, which collapses
to $\mathrm{Flue}(s) = |V_s|$ because the Cosine distance to the zero vector 
is always $\delta(x,0) = 1 - 0 = 1$. As stated in Section~\ref{sec:definition}, this is equivalent to the fluency definition in Torrance's test.
For example, the \enquote{Magnetic Challenge} project contains $|V_s| = 105$ images. This is higher than the average fluency score of the projects we assessed (Table~\ref{tbl:visual-creativity-stat}).

To measure visual flexibility, we use \ref{eq:flex}, where $x$ and $y$ represent the different images in a program, and $|V_s|$ is the total number of these images. For example, the \enquote{Magnetic Challenge} project contains a lot of images, but many of them are visually similar, such as four variants of spikes which only differ in the number of spikes and their orientation. Accordingly, the cosine distance between the corresponding ResNet50 embeddings is very low, which in turn means that they do not contribute much to flexibility. Overall, the flexibility score for \enquote{Magnetic challenge} is $23.29$, which is below the average flexibility score of $27.75$ (Table~\ref{tbl:visual-creativity-stat}), suggesting that the visual flexibility of \enquote{Magnetic challenge} is relatively low.

We compute visual originality as the average pairwise distance between the images in the project and the images in reference projects. As with code, we used three randomly sampled groups of projects for reference. For the \enquote{Magnetic Challenge} project, we thus obtained originality scores of $0.56$ in the first group, $0.55$ in the second, and $0.52$ in the third. The first two scores are close to the mean score of the programs in our study, and the third is the minimum score as shown in Table~\ref{tbl:visual-creativity-stat}, indicating that the visual originality of \enquote{Magnetic challenge} is relatively low.

\begin{table}
\centering
\caption{Visual Creativity Statistics}
\label{tbl:visual-creativity-stat}
\begin{tabular}{cccc}
\toprule
\midrule
{} &  fluency &  flexibility &  originality \\
\midrule
mean& $102.74$ & $27.75$ & $0.57$ \\
std& $106.53$ & $33.69$ & $0.03$ \\
min& $4.00$ & $0.60$ & $0.52$ \\
max& $583.00$ & $192.00$ & $0.66$ \\
\midrule
\bottomrule
\end{tabular}
\end{table}

\subsubsection{Audio creativity}
Projects in Scratch can contain different sounds, either predefined by Scratch, recorded by the user, or uploaded by the user. As with images, the space of possible sounds is infinitely large and varied,
including various file formats, lengths, pitch, etc. Accordingly, we opt for a semantic embedding
approach again.
However, in contrast to images, we do not represent sounds as a single vector, but as a time series
of vectors to take into account how sound can change over time. As a tool for our embedding, we use the pyAudioAnalysis Python package \cite{giannakopoulos2015pyaudioanalysis}. PyAudioAnalysis has been used to extract extract audio features statistics which describe the the wave properties of a sound, including Fourier frequencies, energy-based features, Mel-Frequency Cepstral Coefficients (MFCC), and similar representations~\cite{giannakopoulos2015pyaudioanalysis}.
For each sound we extracted $136$ features. As the sounds differ in sampling rates, we used different window and step sizes: $220$ frames for smaller sampling rates (such as $11\,025$ Hz, $22\,050$ Hz and $24\,000$ Hz) and $250$ for larger sampling rates (such as $44\,100$ Hz and $48\,000$ Hz). Each project is then represented as set of 2-dimensional matrices, each representing a sound. As a distance metric $\delta$, we compute the Euclidean distance between features at matching time stamps as suggested in previous sound comparisons studies~\cite{san2017euclidean}. As the length of the sounds can differ, we apply padding on the end of the shorter representation before calculating the distance as suggested by \cite{le2017agglomerative}. 
Using this distance, we computed audio fluency, flexibility, and originality for 45 Scratch projects. Table~\ref{tbl:statistic_audio} presents the resulting statistics.

Audio fluency is computed using \ref{eq:flu}, where we represent the null concept as a 2-D array filled with zeros. For example, \enquote{Magnetic Challenge} received a fluency score of $1771.69$, which is much lower than the average fluency score of $9160$.

For audio flexibility, we measure the distance between each pair of sounds in the program using \ref{eq:flex}. $x$ and $y$ are different sounds in a program and $|V_s|$ is the total number of these sounds. The \enquote{Magnetic Challenge} program has six sounds ($|V_s| = 6$), and a flexibility score of $1\,177\,416$ which is about a quarter of the mean score of the programs’ in our study as shown in Table~\ref{tbl:statistic_audio}. This indicates that the audio flexibility of \enquote{Magnetic Challenge} is relatively low, as well.

Finally, audio originality is computed as the average distance of each sound in a program to sounds in a sample of reference programs. As with code and visual creativity, we used three different samples, yielding originality scores of $1126.93$, $1204.83$ and $1055.61$ for the \enquote{Magnetic Challenge} project. All three are lower than the mean score of audio originality in our study, indicating that the audio originality of \enquote{Magnetic Challenge} is relatively low.

Overall, the example of \enquote{Magnetic Challenge} indicates that we can pinpoint which modality of a project contributes to creativity (here: the code) and which do not (here: visual and audio). 
This shows that, although the creativity measure of a particular project can be relatively high in one modality, this does not necessarily imply that other modalities in a student's project also exhibit high creativity scores. Hence, it is important to examine them all.

\begin{table}
\centering
\caption{Audio Creativity Statistics}
\label{tbl:statistic_audio}
\begin{tabular}{cccc}
\toprule
\midrule
{} &  fluency &  flexibility &  originality \\
\midrule
mean& $9\,160$ & $44\,771\,043$ & $1744$ \\
std& $12\,744$ & $69\,719\,132$ & $849$ \\
min& $41.20$ & $0$ & $1009$ \\
max& $51\,983$ & $316\,353\,033$ & $4379$ \\
\midrule
\bottomrule
\end{tabular}
\end{table}

\section{Human Creativity Assessment}
\label{sec:web-interface} 

In the previous sections, we established a theory-driven formalization of creativity and
applied it to Scratch programs. In this section, we describe a user study to collect 
expert assessments of creativity in Scratch projects. Each expert was assigned a set of preselected Scratch projects. The experts were Scratch instructors without prior knowledge in creativity theory. Our primary research question is whether we can predict human expert assessments from the automatic creativity measures in Section~\ref{sec:scratch_preprocess}.

We asked the experts to evaluate four different aspects of each project, namely code, visuals, audio, and idea. 
The first three aspects capture the different modalities of Scratch projects, whereas the latter aspect refers to the general idea behind the project. 
As Scratch enables creating different types of projects (stories, tutorials, games, etc.) in diverse domains (math, fashion, medicine, TV shows, etc.), the idea aspect can be important to capture crucial context for an experts' assessment \cite{resnick2009scratch}. 
Additionally, we asked experts to weigh each aspect according to its importance for creativity.
The overall creativity score for each project corresponded to the weighted sum of the creativity
assessments of each aspect.

\subsection{Scratch Creativity Assessment Tool}

\begin{figure}
\includegraphics[width=\linewidth]{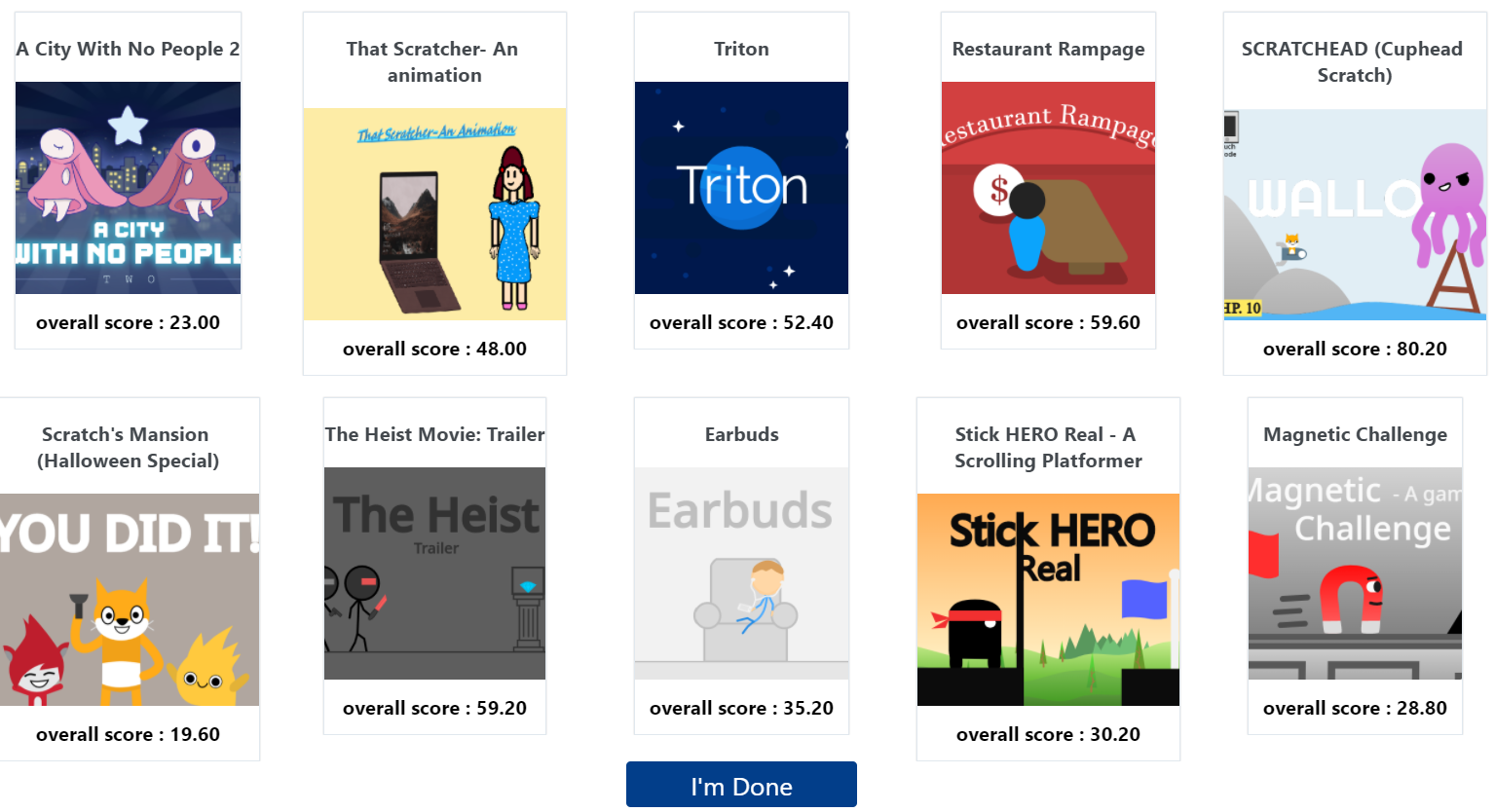}
\caption{Home screen of the assessment tool.}
\label{fig:homescreen}
\end{figure}

To facilitate the assessment process, we developed a web-application named \enquote{Scratch Creativity Assessment Tool.} Experts started on the home screen (Fig.~\ref{fig:homescreen}), which displayed all of the projects that are assigned to an expert. The creativity score of evaluated projects also appeared, allowing experts to compare scores and change them at will. 

\begin{figure*}
\begin{center}
\includegraphics[width=0.3\linewidth]{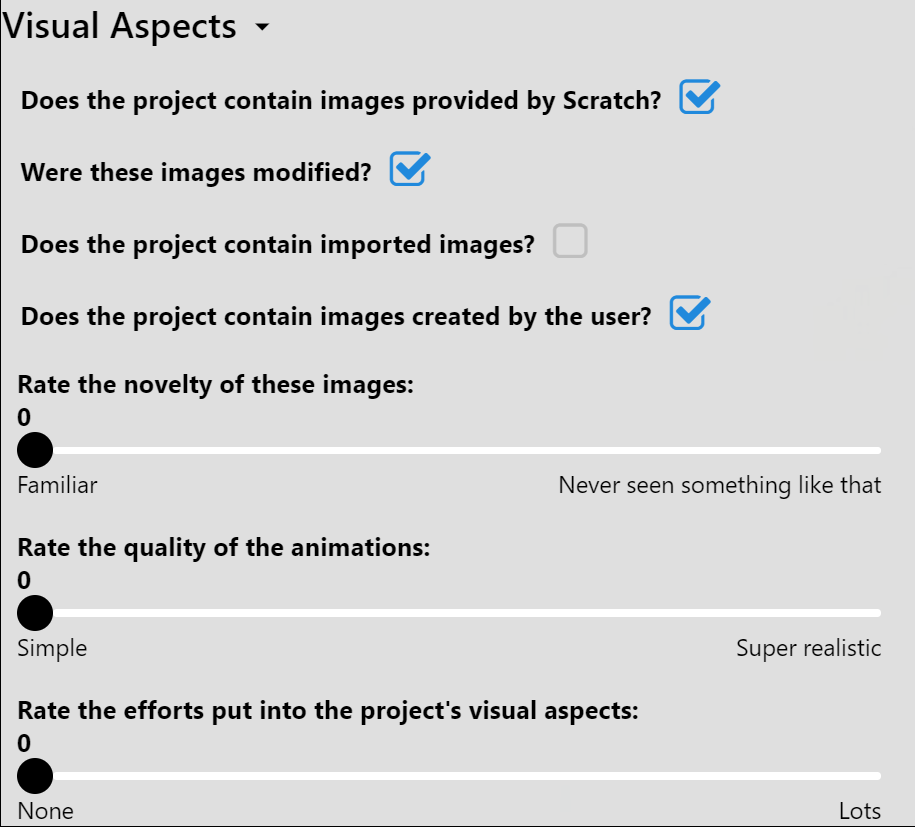}\hspace{0.2cm}
\includegraphics[width=0.3\linewidth]{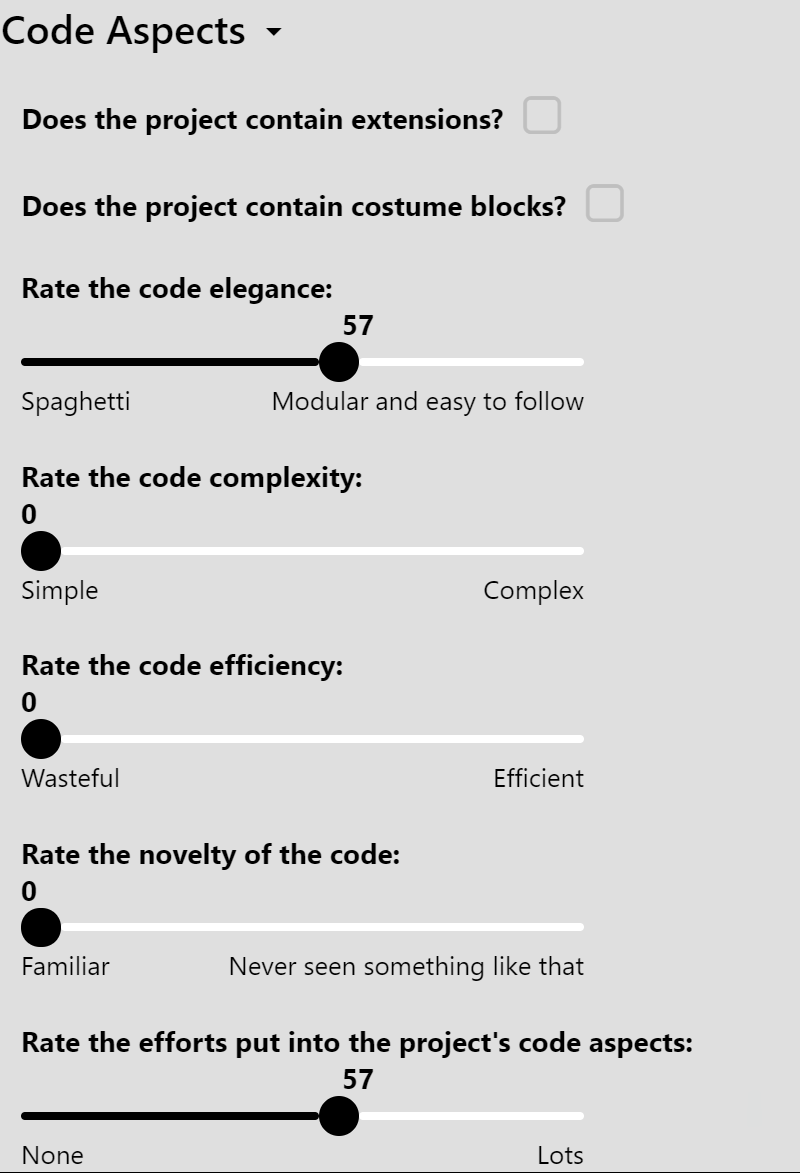}\hspace{0.2cm}
\includegraphics[width=0.3\linewidth]{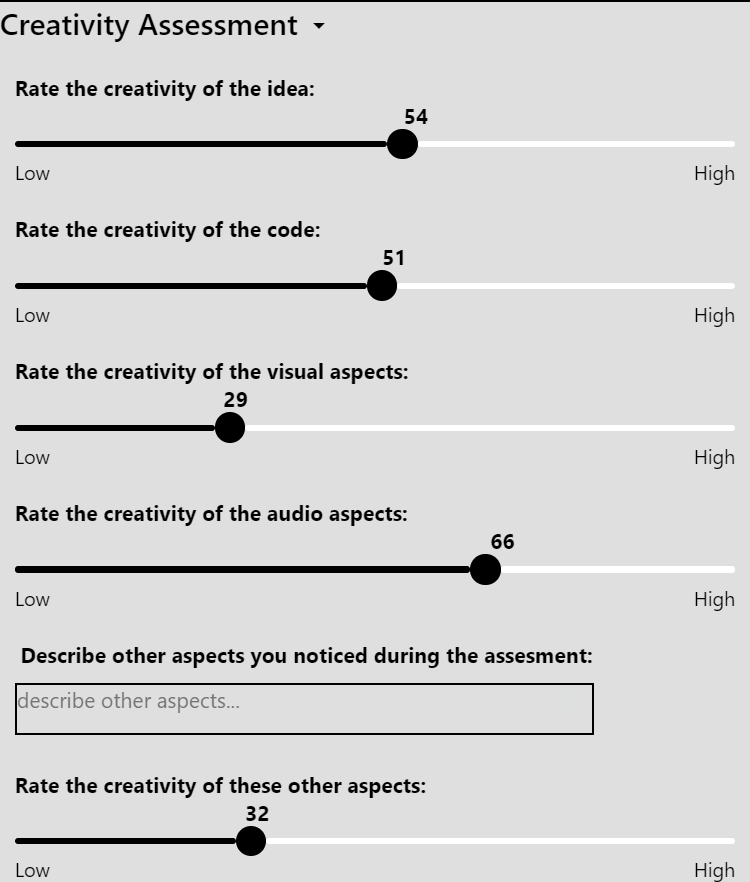}
\end{center}
\caption{Appraisal questions in the user study. (a) Questions on visual creativity. (b) Questions on code creativity. (c) Overall creativity assessment.}
\label{fig:creativityassessment}
\end{figure*}

By clicking on a thumbnail on the home screen, experts could navigate to a creativity questionnaire for the respective project. Each questionnaire began with an interactive display of the project, such that experts could test it before assessing its creativity.
The questionnaire was, then, divided into five section, namely one per aspect (code, visuals, audio, and idea) and a final section for the creativity assessment.
The questions in the first four sections were designed to prime the expert's creativity assessment by presenting them with possible creativity criteria relating to fluency, flexibility, and originality in Scratch projects. For example, the questionnaire regarding the visual aspects (Fig.~\ref{fig:creativityassessment}(a)) contained the binary question \enquote{Does the project contain images created by the user?}, related to fluency. If the answer was \enquote{yes,} then questions about these images were revealed, for example, \enquote{Rate the novelty of these images,} related to originality. An expert could give a rating between 0 and 100 for such questions.

Fig.~\ref{fig:creativityassessment}(b) shows questions relating to the project code, such as evaluating the code complexity, efficiency, and novelty, as well as rating the effort put into the code. Questions about the project idea asked the experts to include a short description of the project's idea as they understand it, and to rate how much novelty and effort were required for developing the idea. If the project included sounds, experts were asked if these sounds were recorded by the user, uploaded, or were predefined by Scratch. Additionally, the experts rated the novelty of the sounds and the effort invested in the audio aspect.

In the final section of the questionnaire (Fig.~\ref{fig:creativityassessment}(c)), we asked experts
to provide creativity scores for each aspect on a scale from 0 to 100. In order to capture aspects 
that went beyond code, visuals, audio, or idea, we included a text field where experts could include
any additional aspect related to creativity, and an additional rating for the creativity of these
additional aspects.

Once experts were done with assessing the creativity of the projects, they returned to the home screen
and clicked on \enquote{I'm done} (Fig.~\ref{fig:homescreen}). This led them to a final screen where
we asked them to weigh the different aspects according to their importance for creativity.
Weights were automatically scaled to values between zero and one, and normalized to a sum up to one.

This concludes our description of the Scratch creativity assessment tool. We now continue to describe our expert recruitment scheme and our study results.

\subsection{Expert Assessment}

For our study, we recruited five experts from four countries: Cuba, Vietnam, India, and Israel. All experts had at least two years experience in teaching Scratch to students of different ages in schools and after-school activities.

We selected 45 unique projects of different types (games and stories), created by different users (age between 9 and 18, from 25 different countries, with 4 to 258 created projects). The assignment of programs to the experts was randomized uniformly. 
Four experts evaluated 20 projects, each, while one evaluated ten projects. $80\%$ of projects were
reviewed by more than one expert. In the remainder of this paper, we designate our experts by the digits one to five.

Table~\ref{tbl:experts-weights} shows the weights assigned by the experts to the different aspects. 
On average, experts assigned the highest weights to idea (range $.25$--$.3$; mean $.29$) and visuals (range $.2$--$.3$; mean $.28$), a medium weight to code (range $.2$--$.3$; mean $.23$), a low weight to audio (range $.1$--$.2$; mean $.15$), and a very low weight to
other aspects (range $.02$--$.1$; mean $.05$). This highlights that a single modality is likely insufficient to capture the full richness of creativity in Scratch projects. On the other hand, the high weight for the idea aspect is a challenge for automatic creativity assessment as the idea of a project is particularly difficult to capture in an automatic computation.

\begin{table}
\centering
\caption{Expert Importance Weights for each aspect}
\label{tbl:experts-weights}
\begin{tabular}{ccccc}
\toprule
\midrule
Expert & Code & Visual & Audio & Idea \\
\midrule
1 & .25 & .35 & .10 & .25  \\
2 & .22 & .28 & .15 & .30  \\
3 & .20 & .20 & .20 & .30  \\
4 & .20 & .30 & .15 & .30 \\
5 & .30 & .25 & .13 & .30 \\
\midrule
mean & .23 & .28 & .15 & .29 \\
\midrule
\bottomrule
\end{tabular}
\end{table}

\begin{table}
\centering
\caption{Expert Creativity Scores}
\label{tbl:experts-stats}
\begin{tabular}{cccccc}
\toprule
\midrule
{Expert} & {Statistic} & Code & Visual & Audio & Final score \\
\midrule
\multirow{2}*{1} & Mean & 69.55 & 75.85 & 45.95 & 67.59 \\
& SD &  24.04 & 24.97 & 23.80 & 21.31 \\
\hline
\multirow{2}*{2} & Mean & 66.75 & 67.70 & 65.80 & 66.89 \\
& SD  & 13.11 & 14.28 & 21.05 & 13.80 \\
\hline
\multirow{2}*{3} & Mean & 70.75 & 77.65 & 59.10 & 65.30 \\
& SD & 10.18 & 10.50 & 29.49 & 11.89 \\
\hline
\multirow{2}*{4} & Mean & 72.90 & 83.40 & 81.80 & 76.11 \\
& SD & 24.00 & 20.11 & 20.06 & 17.78 \\
\hline
\multirow{2}*{5} & Mean & 64.60 & 68.55 & 51.80 & 63.52 \\
& SD & 15.27 & 13.15 & 29.01 & 13.17 \\
\midrule
\bottomrule
\end{tabular}
\end{table}

Table~\ref{tbl:experts-stats} presents the statistics of code, visual, audio, and final creativity scores provided by the experts. We note that not all the projects included audio; therefore, their score for this aspect is zero, which leads to the high standard deviations in this aspect across all experts. We note that the highest scores were provided by expert four and that this expert as well as expert one had the highest standard deviation for visual, code, and final creativity scores. By contrast, experts two and five gave slightly lower scores with smaller standard deviation. 

Overall, we observe that experts vary both in their scores as well as in the importance of each aspect.
In the following section, we analyze the agreement between experts in more detail.

\subsection{Agreement Between Experts}
\label{subsec:experts_agree}

The numeric agreement agreement between experts on overlapping projects was affected by discrepancies in how experts graded the projects assigned to them, as seen in Table~\ref{tbl:experts-stats}.
For example, experts two, four, and five all evaluated the project \enquote{Magnetic Challenge}. Expert two gave this project a creativity score of $82$ for the code aspect, while expert four gave it a score of $42$, and expert five gave a score of $84$. 
Table~\ref{tbl:experts-stats} suggests that experts five and two scored the code aspect on a similar range, whereas expert four gave higher scores with a larger variance.

We note that the differences between experts do not indicate a mistake or lack of expertise. It reflects the fact that creativity assessment is largely subjective \cite{Amabile2018}.
To partially compensate for this, we henceforth measure agreement using Kendall's $\tau$ \cite{abdi2007kendall}, which considers rank agreement rather than numeric agreement. This measure ranges from minus one (complete disagreement between rankings) to plus one (perfect match) and is determined based on overlapping projects for each pair of experts.

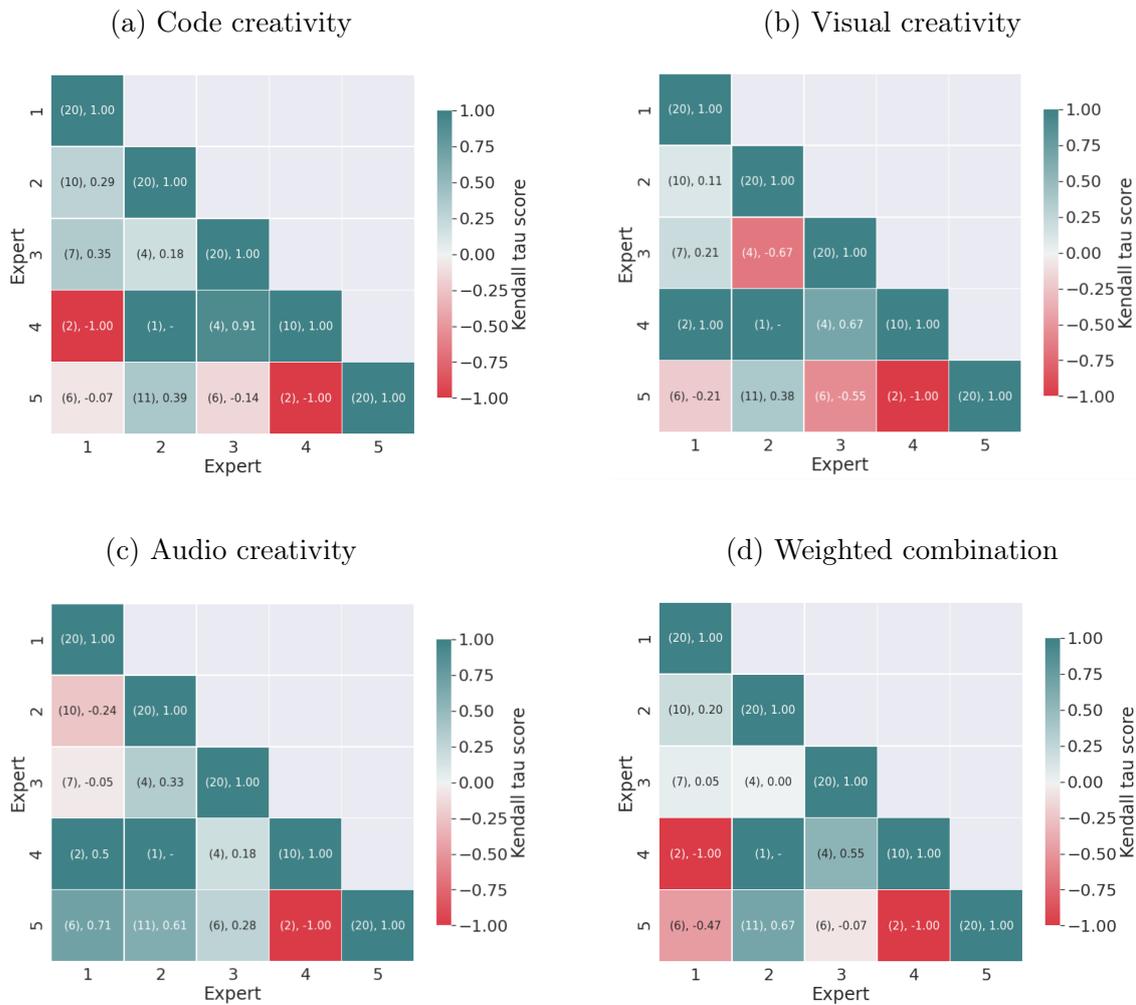
\begin{figure*}
\begin{center}
\begin{tikzpicture}
\node at (0,0) {\includegraphics[width=0.45\columnwidth]{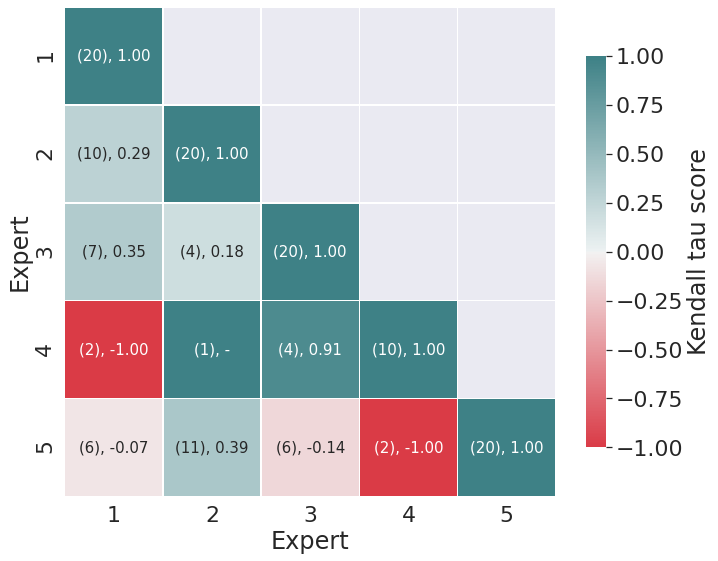}};
\node[above] at (-0.5,3) {(a) Code creativity};
\node at (8,0) {\includegraphics[width=0.45\columnwidth]{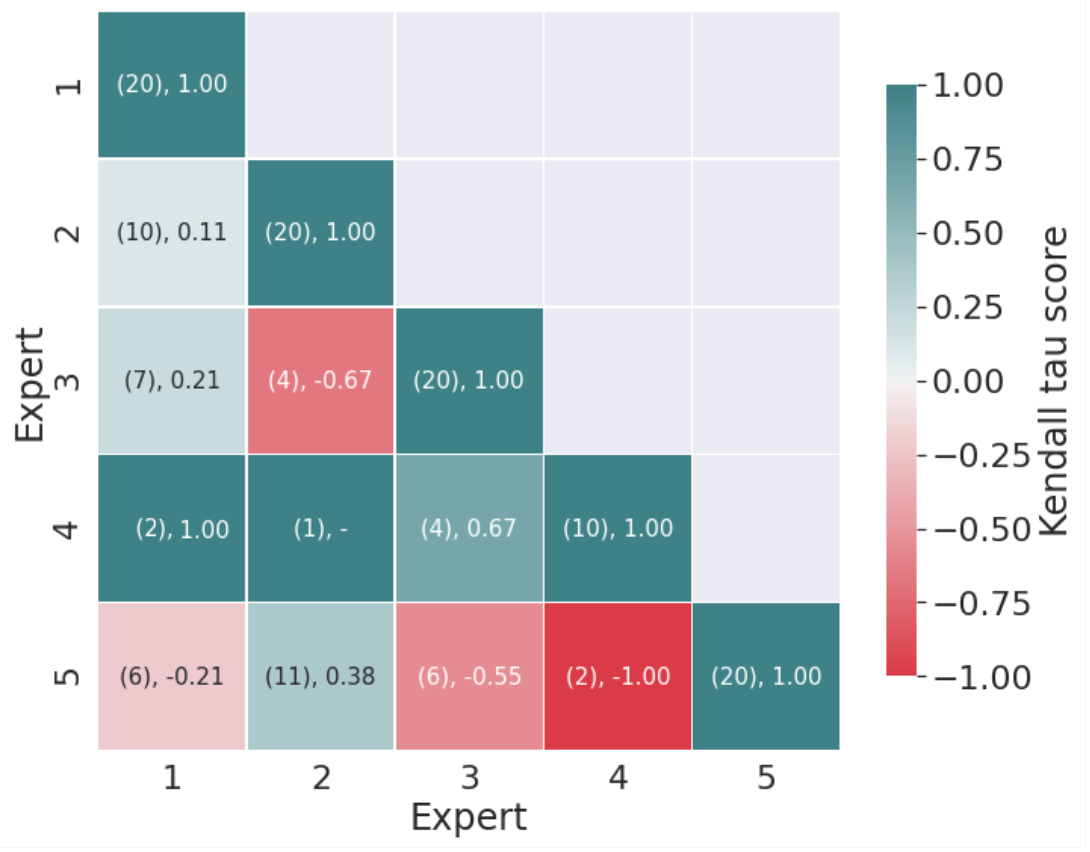}};
\node[above] at (8.2,3) {(b) Visual creativity};
\node at (0,-7) {\includegraphics[width=0.45\columnwidth]{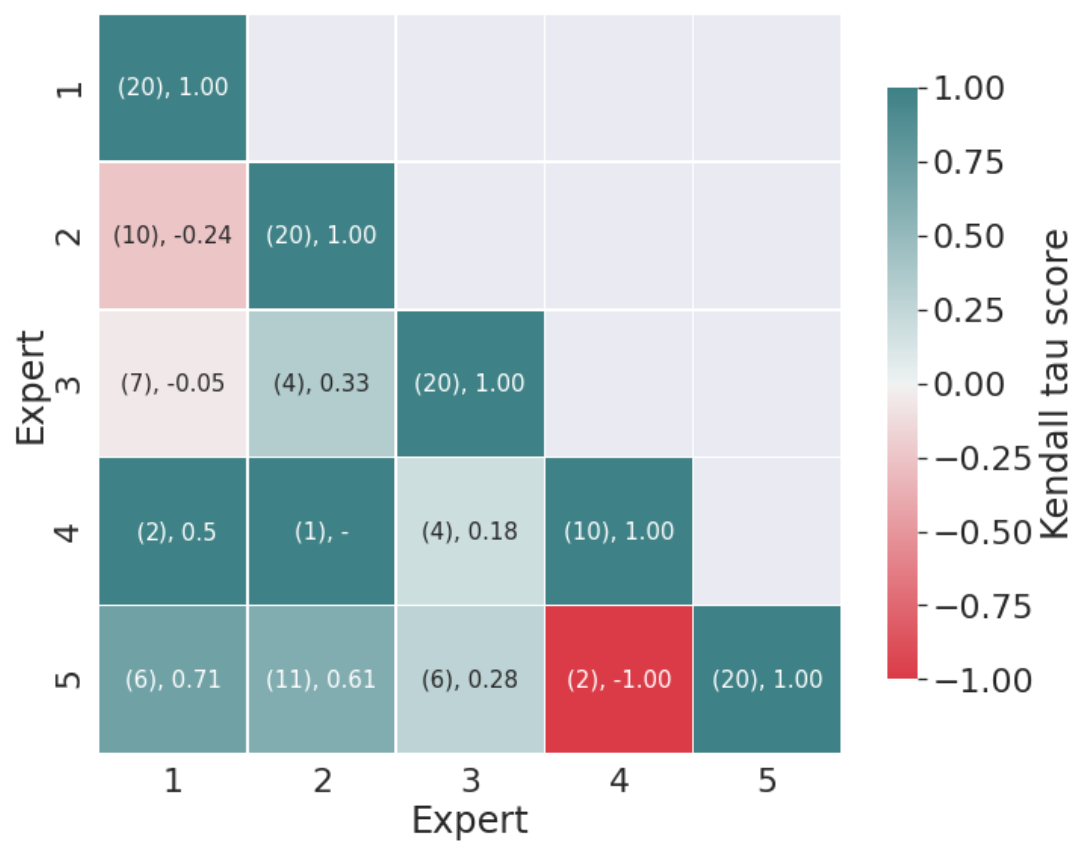}};
\node[above] at (-0.5,-4) {(c) Audio creativity};
\node at (8,-7) {\includegraphics[width=0.45\columnwidth]{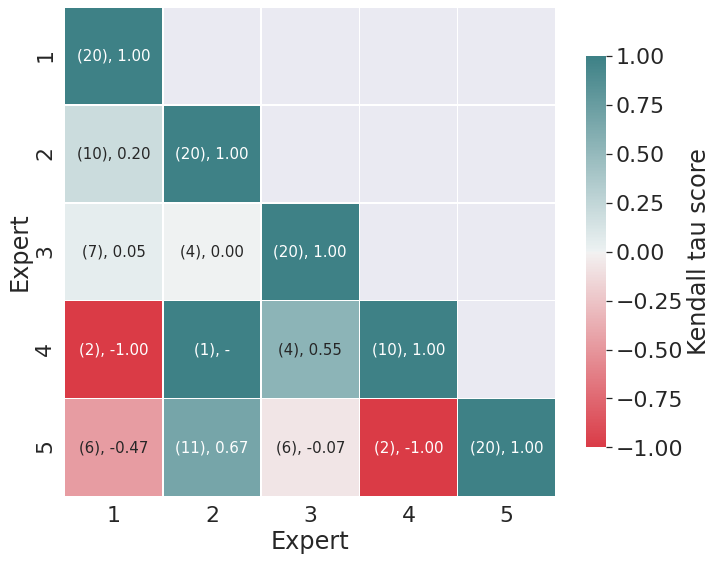}};
\node[above] at (8.2,-4) {(d) Weighted combination};
\end{tikzpicture}
\end{center}
\caption{Kendall's $\tau$ between pairs of experts with overlapping projects. (The number of overlapping projects is shown in parentheses.) (a) Code creativity. (b) Visual creativity. (c) Audio creativity. (d) Weighted sum of visual, code, and audio creativity score.}
\label{fig:agreement_experts_kendall}
\end{figure*}

Fig.~\ref{fig:agreement_experts_kendall} displays the number of overlapping projects (in brackets)
as well as Kendall's $\tau$ for each pair of experts. We evaluated agreement separately for
code (Fig~\ref{fig:agreement_experts_kendall}(a)), visual (Fig~\ref{fig:agreement_experts_kendall}(b)),
and audio creativity (Fig~\ref{fig:agreement_experts_kendall}(c)), as well as the weighted average
of these three scores (Fig~\ref{fig:agreement_experts_kendall}(d)).

Note that experts two and four had only one overlapping project, therefore Kendall's $\tau$ cannot be calculated for them. As shown in the Table, experts one and five mostly agree on audio creativity ($\tau = .71$) but disagree on visual and code creativity as well as for the weighted score. These substantial differences in the rankings of projects in three out of the four scores suggests that they differ in their interpretation of creativity. For example, for the code aspect, the same project was ranked first by expert five and sixteenth by expert one. By contrast, experts two and five have a relatively high agreement for all aspects ($\tau = .38$, $\tau = .39$, and $\tau = .61$, respectively). These experts have relatively similar scores (Table~\ref{tbl:experts-stats}) and weights (Table~\ref{tbl:experts-weights}), suggesting that they interpret creativity in similar ways. 
The highest agreement for visual and code creativity are between experts three and four ($\tau = .67$ and $\tau = .91$). Experts one and five have the highest agreement on the audio creativity, and experts two and five have the highest agreement on the weighted creativity ranking, although they have lower agreement on each aspect separately. This can be explained by the fact that experts two and five
were more consistent in their agreement across aspects. 

Despite some examples of high agreement, the average $\tau$ in Fig.~\ref{fig:agreement_experts_kendall}
is rather low ($-.01$ for code, $-.01$ for visual, $.15$ for audio, and $-.13$ for the weighted combination). We note that this low agreement occurred even though we had primed the experts with the same questionnaire, indicating that even a shared frame of reference is not necessarily sufficient to achieve consistent creativity assessments from human experts. These expert disagreements underline the challenge in quantifying creativity.

Next, we develop a machine learning model to predict human-like creativity assessments from the automatic creativity scores described in Section~\ref{sec:scratch_preprocess}.

\subsection{Predicting Creativity Scores}
\label{sec:predict}

Our aim in this section is to build a machine learning model which receives the automatic creativity
scores of Section~\ref{sec:scratch_preprocess} as input and outputs an overall creativity score for
a Scratch project that is similar to a score a human expert would give. Such a model could serve
as surrogate for a human expert panel and could support students and teachers in assessing the
creativity of Scratch projects swiftly and frequently. We note that we can not expect a high accuracy
of such a model, given that experts disagree with each other, leading to highly noisy training data. 
Nonetheless, we are checking whether a first step in this direction is possible, and whether we can obtain a model that agrees with each expert more than they agree with each other.

We use an XGBoost regressor~\cite{Chen2016} to predict the expert creativity scores for each project. As input features we used the computed originality, flexibility, and fluency measures for visual, code, and audio aspects, as described in Section~\ref{sec:scratch_preprocess}. This provided us with nine features for each project.
The reference sample $S$ for computing originality included all other projects that the each expert rated.

We created two types of XGBoost models: First, a single-expert model trained on projects for each expert separately and, second, a combined model trained on the projects from all experts together. We note that for the combined model, projects that were evaluated by more than one expert were treated as different instances.

For each type of model, we created four different prediction models, 1) predicting the code creativity score, 2) predicting the visual creativity score, 3) predicting the audio creativity score, and 4) predicting the overall project score by the weighted combination score of visual, code, and audio. 
The features consisted of the originality, flexibility, and fluency for the code aspect (model 1), the originality, flexibility, and fluency for the visual aspect (model 2), the originality, flexibility, and fluency for audio aspect (model 3), or all of them (model 4), that is, all nine features (three for originality, three for flexibility, and three for fluency). 

The combined model was evaluated using tenfold crossvalidation. To ensure that the test set contained at least two projects, the single-expert models were evaluated using five folds.

The combined model and the single-expert models were developed using the official implementation of XGBoost \cite{Chen2016}. We selected the hyperparameters based on the structure of our data. In more detail, we set the upper complexity limit of the model to ten trees for the expert with ten projects, fifteen for the rest of the experts, and 29 trees for the combined experts. We set the maximum tree depth based on the number of features.

\begin{table}
\centering
\caption{Kendall's $\tau$ Between XGBoost Regressor and Experts Scores}
\label{tbl:models_kendalltau}
\begin{tabular}{ccccc}
\toprule
\midrule
Expert & Code & Visual & Audio & Weighted combination \\
\midrule
1 &  .52 &  .52 & .49 &  .48 \\
2 &  .51 &  .42 & .32 &  .45 \\
3 &  .53 &  .58 & .69 &  .45 \\
4 &  .46 &  .52 & .43 &  .53 \\
5 &  .52 &  .42 & .44 &  .53 \\
\midrule
Combined & .43 & .44 & .41 & .46 \\
\midrule
\bottomrule
\end{tabular}
\end{table}

We evaluated the creativity score prediction via the following steps. First, we predicted creativity scores for test data projects. Second, we combined these scores with the training data scores to obtain a ranked list of all projects. Finally, we computed Kendall's $\tau$, comparing this ranked list with the expert-given ranked list. This approach was applied using only the pairs with at least one instance from the test data. This method yielded sufficiently many pairs to compute Kendall's $\tau$.

Table~\ref{tbl:models_kendalltau} presents the resulting $\tau$ values.
As seen from the table, when predicting the creativity ranking for visual 
and code aspects, we achieve a $\tau$ of $.51$ and above for three out of five experts. This score is higher than that of the inner-agreement between the experts themselves that is reported in Fig.~\ref{fig:agreement_experts_kendall}(a) and ~\ref{fig:agreement_experts_kendall}(b) (except for the pairs three and four, and one and four). For audio creativity, we achieve a $\tau$ of $.42$ or higher for four out of five experts. This score is higher than the inner-agreement between the majority of experts pairs (except for the three pairs $(1, 5)$, $(1, 4)$, and $(2, 5)$) as shown in Fig.~\ref{fig:agreement_experts_kendall}(c). 
When predicting the weighted creativity score, we achieve a $\tau$ of over $.45$ for all experts. This score is higher than that of the inner-agreement between the majority of experts (except for two pairs $(2, 5)$, and $(3, 4)$´) as suggested by Fig.~\ref{fig:agreement_experts_kendall}(d).  

The bottom row in Table~\ref{tbl:models_kendalltau} presents the results for the combined XGBoost regressor model. Given the disagreement between experts, we would expect that this model performs worse
than the single-expert models. Indeed, this is the case for code creativity. However, for audio creativity, the combined model achieves a higher agreement than the model for expert two,
for visual creativity it achieves a higher agreement than the model for experts two and five, and for
the weighted combination it achieves a higher agreement than the model for experts two and three.
This indicates that the combined model can aggregate data from multiple experts in a useful way,
despite the disagreement between experts.

For all aspects, the combined model achieves $\tau$ above $.41$, which is higher than the agreement scores between most pairs of experts---eight out of nine for the code aspect, six out of nine for the audio aspect, and seven pairs out of nine for the visual aspect as well as the weighed combination---and clearly higher than the average $\tau$ between experts ($-.13$ for the weighted combination).
Overall, our results indicate that an automatic model is able to produce creativity scores which are human-like and
which form a compromise between multiple experts.

\section{Limitations}

While our implementation for Scratch provides first evidence that automatic creativity scores
can approximate human scores, there are still limitations of our study. First, we considered a
rather small sample of experts and our results are not necessarily representative for a broader
group of experts. Second, our machine learning model is trained on the expert's data. For a proper
validation of our model, our scores would have to be compared with a new group of experts.
Third, our fluency model for images and code was rather coarse-grained, especially for custom
code blocks. Future work could improve this model by including more nuanced measures, such as
code complexity \cite{MorenoLeon2016}. Fourth, we used a random reference sample for creativity.
However, we could inject more domain knowledge by adjusting our reference sample, for example
by selecting the demo projects supplied by Scratch or the standard image set provided by Scratch.
Fifth, our current scheme is a static assessment. Future work could investigate how students'
creativity scores evolve over time and how teaching impacts creativity scores.
Finally, any scheme for (automated) assessment raises ethical issues, such as fairness, and ought
to be implemented with care. Future work should investigate how creativity assessment impacts students,
for example, with respect to their confidence and self-image.

\section{Conclusion}

We introduced a novel measure of creativity with three components: the distance to an empty product (fluency), the distance between concepts in the product (flexibility), and the distance to typical products (originality). Our measure only requires two ingredients: a set of concepts that students can use and a (semantic) distance between those concepts. This makes our measure easily applicable across modalities. For example, in this work, we represented code by its syntactic building blocks, we represented images by neural network embeddings, and we represented sound as a sequence of audio features. We showed that our proposed measure is a proper generalization of fluency, flexibility, and originality as proposed by \cite{Torrance1972}.

To validate our distance-based creativity measure, we applied it to Scratch projects and compared it to the creativity assessment of human experts. We found that an XGBoost model using fluency, flexibility, and originality as input agreed at least as well with human assessments as humans agreed with each other. This is partially because human experts tended to disagree, indicating again that creativity is hard to quantify. But it also provides some evidence that automatic measures of creativity can approximate human assessments of creativity.

\section*{Acknowledgment}

This work was partly funded by the German Research Foundation under Grant PA 3460/2-1

\bibliographystyle{plainnat}
\bibliography{literature_shortened}

\end{document}